\documentclass[12pt]{article}
\usepackage{epsfig,psfig,graphicx,rotating}
\usepackage{amsfonts}
\begin{document}
\title{\bf  
STATISTICAL MECHANICS OF THE SELF-GRAVITATING GAS: II. LOCAL PHYSICAL
MAGNITUDES AND FRACTAL STRUCTURES}
\author{{\bf  H. J. de Vega}(a), {\bf N. S\'anchez}(b) \\ \\
%\affiliation{ 
(a)Laboratoire de Physique Th\'eorique et Hautes Energies, \\
Universit\'e Paris VI, Tour 16, 1er \'etage, \\ 4, Place Jussieu
75252 Paris, Cedex 05, FRANCE. \\
Laboratoire Associ\'e au CNRS UMR 7589. \\ 
%\author{\bf N. S\'anchez}
%\affiliation{ 
(b) Observatoire de Paris,  Demirm, \\ 61, Avenue de l'Observatoire, \\
75014 Paris,  FRANCE. \\
Laboratoire Associ\'e au CNRS UA 336, \\
Observatoire de Paris et \'Ecole Normale Sup\'erieure.  }
\date{\today}
\maketitle
\begin{abstract}
We complete our study of the self-gravitating gas by computing
the fluctuations around the saddle point solution for the three
statistical ensembles (grand canonical, canonical and
microcanonical).  Although the saddle point is the same for the three
ensembles, the fluctuations change from one ensemble to the other. The
zeroes of the small fluctuations determinant determine the  position
of the critical points for each  ensemble. This yields the domains of
validity of the mean field approach.  Only the S wave
determinant exhibits critical points. Closed formulae for
the S and P wave determinants of fluctuations are derived.
The {\bf local} properties of the self-gravitating gas in thermodynamic
equilibrium are studied in  detail.
The pressure, energy density, particle density and speed
of sound are computed and analyzed as functions of the position. 
The equation of state turns out to be {\bf
locally } $ p(\vec r) = T \, \rho_V(\vec r) $ as for the ideal gas.  
Starting from the partition function of the self-gravitating gas,
we prove in this microscopic calculation that
the hydrostatic description yielding locally the ideal  gas equation of
state is exact in the $ N = \infty $ limit. The dilute nature of the
thermodynamic limit ($N \sim L \to \infty $ with $N/L$ fixed) together
with the long range nature of the gravitational forces
play a crucial role in obtaining such ideal gas equation.
The self-gravitating gas being inhomogeneous, we have   $ PV/[NT] = f(\eta)
\leq  1 $ for any finite volume $V$.  The
inhomogeneous particle distribution in the ground state suggests a
fractal distribution with Haussdorf dimension $D$, $D$ is slowly decreasing
with increasing density, 
%$ 1 \lesssim D < 3$.
$ 1 < D < 3$.
The average distance between particles is computed in Monte Carlo
simulations and analytically in the mean field approach. A dramatic
drop  at the phase transition is exhibited, clearly
illustrating the properties of the collapse.
\end{abstract}
%\pacs{64.10.+h 04.40.-b 05.45.Df 05.70.Fh}
%\maketitle

\tableofcontents

\section{Statistical Mechanics of  the Self-Gravitating  Gas}

In this paper we continue our investigation of the statistical
mechanics of the self-gravitating gas in thermodynamic equilibrium
initiated in ref.\cite{I}. 
We work in the thermodynamic limit which for the self-gravitating gas
means  the {\bf dilute} limit
\begin{equation}\label{limiT}
N\to \infty\; ,\; V \to \infty\; ,\; {N\over V^{1/3}} = \mbox{fixed}
\end{equation}
where $ V $ stands for the volume of the box containing the gas.
In this limit, the energy $E$, the free energy and the entropy turns to be
extensive. That is, we find that they take the form of $ N $ times a
function of
\begin{equation}\label{defeta}
\eta = {G \, m^2 N \over L \; T} \quad \mbox{or} \quad
\xi = { E \, L \over G \, m^2 \, N^2}
\end{equation}
where $\eta$ and $\xi$ are  intensive variables. Namely, $\eta$ and
$\xi$ stay finite when $ N $ and $ V \equiv L^3 $ tend to infinite.  $\eta$ is
appropriate for the canonical ensemble and $\xi$ for the
microcanonical ensemble. 

In paper I we derive functional integral representations for the
partition function in each of the three statistical ensembles. These
functional integrals are dominated for large $ N $ by a saddle point. 

When any small fluctuation around the saddle point decreases the
statistical weight in the functional integral, the saddle point is
dominating the integral and the mean field approach is fully valid. 
In that case the  determinant of small fluctuations is positive. A
negative determinant of small fluctuations indicates that some
fluctuations around the saddle point are increasing the statistical
weight in the functional integral and hence the saddle point {\bf does
not} dominate the partition function. The mean field approach cannot be
used when the determinant of small fluctuations is negative. 

The zeroes of the small fluctuations determinant determine the  position of the
critical points for  each of the three statistical ensembles. 
The Monte Carlo simulations for the CE and the MCE show that the
self-gravitating gas collapses near the critical points obtained from
mean field. 

The saddle point solution is identical for the three statistical
ensembles. This is not the case for the fluctuations around it. The
presence of constraints in the CE (on the number of particles) and in
the MCE (on the energy and the number of particles) changes the
functional integral over the quadratic fluctuations with respect to
the GCE. 

We compute the determinant of small fluctuations 
around the saddle point solution for spherical symmetry for all three
statistical ensembles. 
In the spherically symmetric case the determinant of small
fluctuations is written as an infinite product over partial waves. The
S and P wave determinants are written in closed form in terms of the
saddle solution. The determinants for higher partial waves are
computed numerically. All partial 
wave determinants are positive definite except for the S-wave. The
S-wave determinant for each ensemble vanishes at the respective critical point.
We find for spherical symmetry: $ \eta^R_{GC} = 0.797375\ldots , \;
\eta^R_{C} = 2.517551\ldots $ and $ \eta^R_{MC} = 2.03085\ldots $ [see
fig. \ref{fig6}]. The variable $ \eta^R $ appropriate for a spherical
symmetry is defined as $ \eta^R \equiv {G \, m^2 N \over R \; T}
= \eta \; \left({4\pi\over 3}\right)^{1/3} =  1.61199\ldots \; \eta\; $ .

The reason why the fluctuations are different in the three ensembles
is rather simple. The more contraints are imposed the smaller becomes
the space of fluctuations. Therefore, in the grand canonical ensemble
(GCE) the system is more free to fluctuate and the phase transition
takes place earlier than in the micro-canonical (MCE) and canonical
ensembles (CE). For the same reason, the transition takes place
earlier in the CE than in the MCE.

The conclusion being that  the MF correctly gives an excellent
description of the thermodynamic limit except near the critical points
(where the small fluctuations determinant vanishes);
the MF is valid for $N|\eta-\eta_{crit}|\gg 1$. The vicinity of the
critical point should be studied in a double scaling limit $N \to
\infty,\; \eta \to \eta_{crit}$. Critical exponents are reported in
paper I for $ \eta \to \eta_C $ using the mean field. These mean field
results apply for $ |\eta-\eta_C| \ll 1 \ll N|\eta-\eta_C| $ and $ N
\gg 1$. Fluctuations around mean field can be neglected in such a regime.

\bigskip

In paper I we expressed all global physical quantities in terms of a
single function  $f(\eta)$. This function obeys a
first order non-linear differential equation of first Abel's type
\cite{kam}.  $f(\eta)$ exhibits a square-root cut at  $\eta_{C}$, the critical
point in the CE. The first Riemann sheet is realized both
in the CE and the MCE whereas the second  Riemann sheet (where $ c_V
< 0 $) is only realized in the MCE.

Besides the global physical quantities computed in paper I, we compute
here the local energy density $ \epsilon(r) $, the local particle
density, the local pressure and the local speed of sound. 

The particle distribution $\rho_V({\vec q})$ proves to be {\bf inhomogeneous}
(except for $\eta \ll 1$) and described by an universal function of $\eta$,
the geometry and the ratio ${\vec r} = {\vec q} / R, \; R$ being the
radial size. Both Monte Carlo simulations and the Mean Field approach
show that the 
system is inhomogeneous forming a clump of size smaller than the box
of volume $ V $ [see figs. 3 and 5 in paper I]. The particle density in
the bulk behaves as $\rho_V({\vec q}) \simeq r^{D-3}$. 
That is, the mass $ M(R) 
$ enclosed on a region of size $ R $ vary approximately as
$$
 M(R) \simeq C \;  R^D \; . 
$$
$ D $ slowly decreases from the value $ D = 3 $ for the ideal gas
($\eta=0$) till $ D = 0.98 $ in the extreme limit of the MC point
taking the value $1.6$ at $\eta_C$ [see Table 2]. 
This indicates the presence of a fractal distribution with
Haussdorf dimension $D$. 

\bigskip

Our study of the statistical mechanics of a self-gravitating system indicates 
that gravity provides a dynamical mechanism to produce fractal
structures\cite{natu}-\cite{gal}.

\bigskip

The average distance between particles monotonically decrease with $
\eta $ in the first sheet. The mean field and Monte Carlo are very
close in the gaseous phase whereas the Monte Carlo simulations exhibit
a spectacular drop in the average particle distance at the clumping
transition point T (see fig.\ref{rmmcmf} ). In the second sheet (only
described by the MCE) the average particle distance increases with $
\eta $ (see fig.\ref{fig15}). 

We find that the {\bf local} equation of state is given by $ p(\vec r) = T \,
\rho_V(\vec r) $. We have thus {\bf derived} the equation of state for the
self-gravitating gas. It is {\bf locally} the {\bf ideal gas
} equation, but the self-gravitating gas being inhomogeneous,  the
pressure at the surface of a given volume is not equal
to the temperature times the average density of particles in the volume.
In particular, for the whole volume: $ PV/[NT] = f(\eta) \leq 1 $
(the equality holds only  for $
\eta = 0 $).  

Notice that we have found 
the local ideal gas equation of state $ p(\vec r) = T \, \rho_V(\vec r) $ 
for purely gravitational interaction between
particles. Therefore, non-ideal gas equations of state (as often
assumed and used in the literature\cite{chandra,sas,pad,bt}) imply the
presence of additional non-gravitational forces. Such non-ideal
equations of state appear for quantum gases \cite{llms}.

\bigskip

The energy density turns out to be an increasing function of $r$ in the
spherically symmetric case. The energy density is always positive on
the surface whereas it is positive at the center for $ 0 \leq \eta^R <
\eta_3^R = 1.73745\ldots $ and 
negative beyond the point $ \eta^R = \eta_3^R = 1.73745\ldots $.

The speed of sound is computed in the mean field approach as a
function of the position for spherical symmetry and long wavelengths. 
$ v_s^2(r) $ diverges at $ \eta^R =  \eta_0^R = 2.43450\ldots $ in the
first Riemann sheet. Just beyond this point $ v_s^2(r) $ is large and
{\bf negative} in the bulk showing the strongly unstable behaviour of
the gas for such range of values of $ \eta^R $. 

\bigskip

Moreover, we have
shown the equivalence between the statistical mechanical
treatment in the mean field approach and the hydrostatic description
of the self-gravitating gas \cite{sas}-\cite{bt}.

The success of the hydrodynamical description depends on the value of
the mean free path ($l$) compared with the relevant sizes in the system. 
$ l $ must be $ \ll 1 $. 
We compute   the ratio $ l/a $ (Knudsen number) where $ a $ is a
length scale that stays fixed for $ N \to \infty $
and show that  $ l/a \sim N^{-2} $. This
result ensures the accuracy of the hydrodynamical description for
large $ N $. 

Furthermore, we have computed in this paper several
physical magnitudes as functions of $ \eta $ and $ r $ which were not
previously 
computed in the literature as the speed of sound, the energy density,
the average distance between particles and we notice the presence of a
Haussdorf dimension in the particle distribution. 

\bigskip

This paper is organized as follows. In section II
we summarize the main results from \cite{I} on the mean field approach,
in sec. III we compute the small fluctuations around the
mean field theory solution. Sec. IV presents our results for space
dependent quantities as the particle distribution, the average
distance between particles, the local energy density, the local
pressure and the local speed of sound. In section V we discuss the
generalization for a space with 
any number of dimensions. In sec. VI the molecular clouds in the
interstellar medium are discussed as a self-gravitating gas.
Discussion and remarks are presented in section VII whereas
appendix A-B contain relevant mathematical developments. 

\section{Summary of the Mean Field Results}

Let us summarize here the main results of paper I in the mean field
approach for spherical symmetry which will be used in what follows.

The saddle point is given by
\begin{equation}\label{fixi}
\phi(r) = \log \rho(r) = \log\left({\lambda^2 \over  4\, \pi \,\eta^R}\right) +
\chi(\lambda \, r) \; .
\end{equation}
Here $ \rho(r) $ is the particle density and $ \chi(\lambda ) $ obeys
the equation
\begin{equation}\label{ecuaxi}
\chi''(\lambda) + {2 \over \lambda} \, \chi'(\lambda) +
e^{\chi(\lambda)} = 0 \quad , \quad \chi'(0) = 0 \quad , \quad
\chi(0) = 0\ ; .
\end{equation}
$ \chi(x) $ is independent of $ \eta^R $, and  $ \lambda $ is related to
 $\eta^R$ through 
\begin{equation}\label{lambaxi}
\lambda \; \chi'(\lambda ) = -\eta^R \; .
\end{equation}
We have in addition,
\begin{equation}\label{fide1}
\phi(1) = \log\left[ {3 \, f_{MF}(\eta^R) \over 4 \, \pi }\right]
\quad , \quad \rho(1) = {3 \over 4 \, \pi } \; f_{MF}(\eta^R) \; .
\end{equation}
where 
\begin{equation}\label{fetexpl}
f_{MF}(\eta^R) = {\lambda^2 \over 3 \, \eta^R}\; e^{\chi(\lambda)}
\quad , 
\end{equation}
The function $ f_{MF}(\eta^R) $ obeys the Abel equation,
\begin{equation}\label{abel}
\eta^R(3f_{MF}-1)f'_{MF}(\eta^R)+(3f_{MF}-3+\eta^R) f_{MF} = 0 \;.
\end{equation}
For $ \eta^R = 0 $ it follows from eq.(\ref{abel}) that
\begin{equation}\label{eta0}
 f_{MF}(0) = 1 \; .
\end{equation}

For the main physical magnitudes in the mean field approach,
[that is, from the above saddle point and neglecting the fluctuations
around it] we find:
\begin{eqnarray}\label{abel2}
{p V \over NT} &=& f_{MF}(\eta^R) \cr \cr
{F - F_0 \over NT}  &=& 3 [1 - f_{MF}(\eta^R)] - \eta^R + \log f_{MF}(\eta^R)
\cr \cr
{S - S_0 \over N} &=& 6 [f_{MF}(\eta^R)-1] + \eta^R - \log f_{MF}(\eta^R)
\\ \cr
{E \over NT} &=& 3 [f_{MF}(\eta^R)-\frac12] \; , \nonumber 
\end{eqnarray}

We find for the speed of sound squared at the surface 
\begin{equation}\label{vsmf}
{v_s^2 \over T} = {  f_{MF}(\eta^R) \over 3 }\, \left[ 4 + { 3\,
f_{MF}(\eta^R) + \frac{\eta^R}2 - 2 \over 6 \, f_{MF}^2(\eta^R)+
\left(\eta^R - \frac{11}2 \right) f_{MF}(\eta^R) + \frac12 } \right]  \;,
\end{equation}
The specific heat at constant volume takes the form
\begin{equation}\label{cvmf}
(c_V)_{MF}= 6 \, f_{MF}(\eta^R)- \frac72 + \eta^R + {\eta^R - 2 \over
3 \, \, f_{MF}(\eta^R) - 1} \; ,
\end{equation}
whereas for the specific heat at constant pressure we find 
\begin{equation}\label{cpmf}
(c_P)_{MF}= 12 \, f_{MF}(\eta^R)- \frac32 + {24\left(\eta^R -
2\right) f_{MF}(\eta^R) \over 6 \, \, f_{MF}(\eta^R) - \eta^R}
\end{equation}

\section{Calculation of the Functional Determinants: the Validity of
Mean Field} 

The mean field gives the dominant behaviour for $ N \to \infty $. We
evaluate in this section the Gaussian functional integral of small
fluctuations around the stationary points. 

As remarked in paper I, the three statistical ensembles (grand
canonical, canonical and microcanonical) yield identical results for
the saddle point. However, the small fluctuations around the
saddle take different forms in each ensemble.

Let us recall the partition functions for the three ensembles keeping
quadratic fluctuations around the saddle point. 
In the grand canonical ensemble (see eq.(VI.29) in paper I), we have
\begin{eqnarray}\label{gausgc}
&&{\cal Z}_{GC}(z,T) = e^{{N\over  4\pi\eta }\; \left\{
\int_0^1 d^3x \left[ \frac12\phi \; \nabla^2_r\phi \; + 4\pi\eta \;
e^{\phi({\vec x})}\right] - 2 \pi\eta \log C(\eta) \right\} } \times
\cr \cr
&&\int\int\;
{\cal D}Y\;e^{ {N \over 8\pi\eta}
\;\int_0^1 d^3x \; \left[ Y \nabla^2 Y +  4\pi\eta \, Y^2
\;e^{\phi({\vec x})} \right]}\left[ 1 + {\cal O}\left( {1 \over
N}\right) \right]   
\end{eqnarray}

\bigskip

In the canonical ensemble, we have for the coordinate partition function,
\begin{equation}\label{gaussy}
e^{\Phi_N(\eta)} \buildrel{ N>>1}\over= e^{-N\,s(\eta)} \; \int\int DY
\, dy_0 \; e^{-N \, s^{(2)}_C[Y(.),y_0] } \; 
\left[ 1 + {\cal O}\left( {1 \over N}\right) \right] 
\end{equation}
where
\begin{equation}\label{scuad}
s^{(2)}_C[Y(.),y_0] = \frac12 \; \int  d^3x  \; { Y^2({\vec x}) \over
\rho({\vec x})} - {\eta \over 2} \int  {d^3x  \; d^3y \over |{\vec
x}-{\vec y}|}\;  Y({\vec x})\;  Y({\vec y}) - y_0 \; \int  d^3x \; Y({\vec x})
\end{equation}
and
$$
s(\eta^R)  =   3[1 - f_{MF}(\eta^R)]- \eta^R+ \log\left[ {3
f_{MF}(\eta^R)\over 4\pi}\right] \; .
$$

\bigskip

In the microcanonical ensemble, we have for the coordinate partition function,
\begin{equation}\label{gausmic}
w(\xi,N) \buildrel{ N>>1}\over=e^{-N\,s(\eta)} \; \int\int DY
\, dy_0 \;{ d {\tilde \eta} \over 2 \pi i}\; e^{-N \,
s^{(2)}_{MC}[Y(.),y_0,{\tilde \eta}] } \;  \left[ 1 + {\cal O}\left( {1 \over
N}\right) \right]  
\end{equation}
where, 
\begin{equation}\label{s2mc}
s^{(2)}_{MC}[Y(.),y_0,{\tilde \eta}] = s^{(2)}_{C}[Y(.),y_0]- {\tilde
\eta} \int  {d^3x  \; d^3y \over |{\vec
x}-{\vec y}|} \;\rho_s({\vec x}) \; Y({\vec y})- {3 \over 4 \,
\eta_s^2}\, {\tilde \eta}^2 \; .
\end{equation}

Fluctuations of order higher than quadratic contribute to the $ 1/N $
corrections in the three ensembles.  

As remarked in paper I,
the functional integral for the canonical and microcanonical
ensembles, eqs.(\ref{gaussy}) and (\ref{gausmic}), respectively,
are rather close; in the last one, eq.(\ref{gausmic}),  there is an
extra integration over one variable that constrains the energy. 

\bigskip

In the grand canonical functional integral (\ref{gausgc}), we have to
compute the determinant of the operator
$$
L_{\rho}(\vec x,\vec y ) \equiv \delta(\vec x - \vec y)\left[ -\nabla^2 -
4\pi \, \eta  \; e^{\phi({\vec x})} \right] \; .  
$$

In the canonical functional integral (\ref{scuad}), we find the operator 
$$
K_{\rho}(\vec x,\vec y) \equiv {\delta(\vec x - \vec y) \over
e^{\phi({\vec x})}} - {\eta \over | \vec x - \vec y |}  
$$
These operators are related by 
\begin{equation}\label{dificil}
-\nabla^2_{\vec x} [  K_{\rho}(\vec x,\vec y)\; e^{\phi({\vec y})}] =
 L_{\rho}(\vec  x,\vec y ) \; .
\end{equation}
Since,
$$
L_0(\vec x,\vec y ) = -\delta(\vec x - \vec y) \, \nabla^2 \quad ,
\quad K_0(\vec x,\vec y ) = \delta(\vec x - \vec y) \; ,
$$
eq.(\ref{dificil}) can be written in abstract form as,
\begin{equation}\label{abs}
L_0 \; K_{\rho} \; D = L_{\rho}; .
\end{equation}
Here,
$$
D(\vec x,\vec y ) = \delta(\vec x - \vec y)\; e^{\phi({\vec y})} \; .
$$
Therefore, taking the determinant of both sides of eq.(\ref{abs})
yields, 
\begin{equation}\label{relKL}
\mbox{Det}\left( { K_{\rho} \over K_0 }\right) = \mbox{Det}\left( {
L_{\rho} \over L_0 }\right) \; e^{ - \int d^3x \; \phi({\vec x})} \; ,
\end{equation}
where we used that for any trace class operator $ M $ as $ D $ 
$$
\log \mbox{Det} M = \mbox{Tr} \log M \; .
$$
That is,
$$
\mbox{Det} D =  e^{ \int d^3x \; \phi({\vec x})}
$$
The  functional determinants are well defined in eq.(\ref{relKL}) by
normalizing with respect to the vacuum values at zero density $  \rho
= e^{\phi} = 0 $.

For spherically symmetric stationary points $ \phi(r) $ we can expand
the determinants in partial waves:
$$
\log \mbox{Det} \left( { L_{\rho} \over L_0 }\right) =
\sum_{l=0}^{\infty} (2l+1) \, \log \mbox{Det}\left({L^l_{\rho}
\over  L^l_0 }\right) \; .
$$
where
$$
L^l_{\rho} = \delta(r-r')  \left[{d^2\over dr^2} + \frac2{r} {d
\over dr}- {l(l+1) \over r^2}  +  4\pi \eta^R \; e^{\phi(r)} \right]
$$
We evaluate these partial wave functional determinants in Appendix A.

We expand the fluctuations in partial waves,
\begin{equation}\label{armesf}
Y(\vec x) = \sum_{l,m} c_{l,m} \; y_l(r) \; Y_{l,m} ({\check r}) \; ,
\end{equation}
where the $ Y_{l,m} ({\check r}) $ are spherical harmonics and the $
c_{l,m} $ arbitrary coefficients.

The small fluctuations determinants explicitly depend on the
boundary conditions imposed to the fluctuations $ Y(\vec x) $ around
the mean field stationary point. 

Following the arguments by Hurwitz and Katz\cite{HK}, one assumes
that outside the sphere (no sources) the fluctuations obey the Laplace
equation and therefore 
$$
y_l(r) = { A \over r^{l+1}} \quad \mbox{for} \; r> 1 \; .
$$
Here $ A $ is some constant. Therefore, imposing continuity for $
Y(\vec x) $ and its radial derivative at $ r = 1 $ yields,
\begin{equation}\label{condfis}
0 = {d \over dr}\left[ r^{l+1} \, y_l(r) \right] = y_l'(1) +
(l+1)\,y_l(1) =  0 \; ,
\end{equation}
We impose this condition to the solutions in Appendix A.

\subsection{The Grand Canonical Ensemble}

Evaluating the Gaussian functional integral in eq.(\ref{gausgc}) yields
\begin{equation}\label{silla}
{\cal Z}_{GC}(z,T)\buildrel{ N>>1}\over= {e^{-N\,s_{GC}(\eta^R)} \over
\sqrt{\mbox{Det}_{GC}(\eta^R)}} \left[ 1 + {T(\eta^R)\over N} + {\cal
O}\left({1 \over N^2}\right)\right] 
\end{equation}
where $ s_{GC}(\eta^R) $ stands for the `effective action' at the
saddle point 
$$
s_{GC}(\eta^R)= \frac12 \, K(\eta)-1 = 2 - 3 \,  f_{MF}(\eta^R) 
$$
where we used eqs.(VI.17) and (VI.66) from paper I.

Det$_{GC}(\eta^R)$ stands for the determinant of small fluctuations
around this spherically symmetric saddle point  and
$ T(\eta^R) $  for the two-loop corrections. Det$_{GC}(\eta^R)$ can be
expressed as an infinite product over the partial waves 
$$
\mbox{Det}_{GC}(\eta^R) =  \mbox{Det}\left( {
L_{\rho} \over L_0 }\right) = \prod_{l\geq 0} [\Delta_l (\eta^R)]^{2l+1} \, .
$$
where $ \Delta_l (\eta^R) = \mbox{Det} \left({L_{\rho}^l \over L_0^l }
\right) $. This infinite product can be properly defined using, for
example, dimensional regularization\cite{adv}. One finds in this way
that it takes a finite value.

$ \Delta_0(\eta^R)$ and  $\Delta_1(\eta^R)$ can be
computed in closed form [see eqs.(\ref{detS}) and (\ref{detP})]
\begin{equation}\label{detSP}
\Delta_0(\eta^R) = 1 - \frac32 \, \eta^R \, f_{MF}(\eta^R) \quad , \quad 
\Delta_1(\eta^R) = { 3 \, \eta^R \over \lambda^2}f_{MF}(\eta^R) \; .
\end{equation}
We see that $ \Delta_0(\eta^R) > 0 $ for $ 0 \leq \eta^R < \eta^R_{GC} =
0.797375\ldots $. Therefore, the mean field approach breaks down for
the grand canonical ensemble at $ \eta^R_{GC} $. 

Notice that the determinants for all waves except the S-wave are
positive definite for $ \eta^R_C > \eta^R \geq 0 $. 

Adding the contributions from the functional determinant to the mean
field results eq.(\ref{abel2}) in the grand canonical ensemble yields,
\begin{eqnarray}\label{cfluc}
{F - F_0 \over NT}  &=& 3 [1 - f_{MF}(\eta^R)] - \eta^R + \log
f_{MF}(\eta^R)
+ {1 \over 2N} \log\mbox{Det}_{GC}(\eta^R)+{\cal O}\left({1 \over N^2}\right)
\cr \cr
{p V \over NT} &=& f_{MF}(\eta^R) +{\eta^R \over 6N}{d \over
d\eta^R}\log\mbox{Det}_{GC}(\eta^R) +{\cal O}\left({1 \over N^2}\right)
\cr \cr
{S - S_0 \over N} &=& 6 [f_{MF}(\eta^R)-1] + \eta^R - \log
f_{MF}(\eta^R)+ {1 \over 2N}\left( \eta^R{d \over d\eta^R} -1
\right)\log\mbox{Det}_{GC}(\eta^R) +{\cal O}\left({1 \over N^2}\right)
\cr \cr
{E \over NT} &=& 3 [f_{MF}(\eta^R)-\frac12]+  {\eta^R \over 2N}\; {d
\over d\eta^R} \log\mbox{Det}_{GC}(\eta^R)+{\cal O}\left({1 \over
N^2}\right) \; ,
\end{eqnarray}
where we used eqs.(VI.16), (VI.20) and  (VI.23) from paper I.

These results correspond to include $ 1/N $ corrections in the
function $ f(\eta^R) $ as follows,
$$
f(\eta^R) = f_{MF}(\eta^R) + {\eta^R \over 6N}{d \over
d\eta^R}\log\mbox{Det}_{GC}(\eta^R) +{\cal O}\left({1 \over
N^2}\right) \; .
$$
Eqs.(\ref{abel2})-(\ref{cpmf}) permit to compute the various physical
quantities in terms of $ f(\eta^R) $. 

Since,
$$
\eta^R{d \over d\eta^R}\log\mbox{Det}_{GC}(\eta^R) \buildrel{ \eta^R \uparrow
\eta^R_{GC}}\over= -{\eta^R_{GC} \over \eta^R_{GC}-\eta^R} \to -\infty \; ,
$$  
$ {p V \over NT} $, the energy and the entropy tend to {\bf minus
infinity} when $ \eta^R \uparrow \eta^R_{GC} $. This behaviour
correctly suggests that the  gas collapses for $ \eta^R \uparrow
\eta^R_{GC} $. Indeed, the Monte Carlo simulations yield a large and
negative value for $ {p V \over NT} $ in the collapsed phase
(paper I). 

We want to stress that the mean field values provide {\bf excellent}
approximations as long as $ N|\eta^R_{GC}-\eta^R| >> 1 $ in the grand
canonical ensemble. Namely, the mean field is completely
reliable for large $ N $ unless $\eta^R$ gets at a distance of the
order $ N^{-1} $ from  $\eta^R_{GC}$.

\subsection{The Canonical Ensemble}

We have to compute the Gaussian functional integral in eq.(\ref{gaussy})
\begin{equation}\label{gaucano}
\int\int DY \, dy_0 \; e^{-N \, s^{(2)}_C[Y(.),y_0] } \; .
\end{equation}
where $ s^{(2)}_C[Y(.),y_0] $ is given by eq.(\ref{scuad}). The
simplest way is to find a saddle point for $ Y(.) $ in
eq.(\ref{gaucano}), that is, a solution $ {\tilde Y}(\vec x) $ of the
equation 
\begin{equation}\label{gauens}
{{\tilde Y}(\vec x) \over \rho({\vec x})} - \eta \int{ d^3y \;
{\tilde Y}({\vec y})\over |{\vec x}-{\vec y}|} - y_0 =0 \; .
\end{equation}
It is convenient to write such solution as $ {\tilde Y}(\vec x) = y_0
\, \rho({\vec x}) \, w({\vec x}) $ and shift the integration variable
in eq.(\ref{gaucano}) as follows
\begin{equation}\label{camvar2}
Y(\vec x) = \rho({\vec x}) \left[y_0\, w({\vec x})+ Z({\vec x}) \right]
\end{equation}
where $ Z({\vec x}) $ is the new functional integration variable and $
w({\vec x}) $ is a solution of the equation
\begin{equation}\label{ecw}
w({\vec x}) - \eta \int{ d^3y \;
\rho({\vec y})\; w({\vec y})\over |{\vec x}-{\vec y}|} - 1 =0 \; .
 \end{equation}
$ s^{(2)}_C[Y(.),y_0] $ takes now the form
$$
s^{(2)}_C[Y(.),y_0] = \frac12 \; [y_0]^2 \; \int d^3x \, \rho({\vec
x}) \; w({\vec x}) + A[Z(.)] \; ,
$$
where
\begin{equation}\label{adeZ}
 A[Z(.)] = \frac12 \int d^3x \; \rho({\vec x})\; Z^2({\vec x})-
{\eta \over 2} \int{ d^3x \; d^3y \over |{\vec x}-{\vec y}|}
\;\rho({\vec x})\;\rho({\vec y})\; Z({\vec x})\; Z({\vec y}) \; .
\end{equation}
We then have,
\begin{eqnarray}\label{resucano}
\int\int DY \, dy_0 \; e^{-N \, s^{(2)}_C[Y(.),y_0] }&=& J \; \int DZ\,
dy_0 \; e^{-{N \over2}\; [y_0]^2 \; \int d^3x \, \rho({\vec x}) \;
w({\vec x})}\; e^{- N \, A[Z(.)]} \cr \cr
&=& { 1 \over
\sqrt{\mbox{Det}\left({K_{\rho} \over K_0}\right) \, 
\int d^3x \, \rho({\vec x}) \; w({\vec x})}} \; ,
\end{eqnarray}
where we used eq.(\ref{relKL}) and  the jacobian $ J $ of the
change of variables (\ref{camvar2})  has the value
\begin{equation}\label{jaco}
J = e^{ \int d^3x \; \phi(\vec x) } \; .
\end{equation}

In the spherically symmetric case eq.(\ref{gauens}) has a spherically symmetric
solution  $ w(r) $ which can be expressed in terms of the stationary
point solution $ \phi(r) $ as follows,
$$
w(r) = {1 \over 2 - 3 \, \eta^R \, f_{MF}(\eta^R)} \left[ 2 + r {d
\phi \over dr}\right]
$$
 [$ w(r) $ is related to the S-wave regular solution (\ref{ondaS})]. We
 can then compute the integral in the r. h. s. of eq.(\ref{resucano})
 with the result,
$$
\int d^3x \,  e^{\phi(r)} \; w(r) = {3 \, f_{MF}(\eta^R)  - 1 \over 2-3
\, \eta^R \, f_{MF}(\eta^R)} 
$$
The argument of the square-root in eq.(\ref{resucano}) becomes then
\begin{equation}\label{DCeta}
D_C(\eta^R) \equiv \mbox{Det}\left({K_{\rho} \over K_0}\right) \, \int d^3x \,
\rho({\vec x}) \; w({\vec x}) =  \frac12 \left[ 3 \,  f_{MF}(\eta^R) - 1
\right] \; \prod_{l\geq 1} [\Delta_l(\eta^R)]^{2l+1} \;
e^{-\int d^3x \; \phi(\vec x) } \;  .
\end{equation}
where we used eqs.(\ref{relKL}), (\ref{detSP}) as well as
$$ 
\Delta_0(\eta^R) \;  \int d^3x \,  e^{\phi(r)} \; w(r) =
\frac12 \left[ 3 \,  f_{MF}(\eta^R) - 1 \right]
$$
and we normalize to unit at $ \eta^R = 0 $. All factors in eq.(\ref{DCeta}) are
 positive definite except the first one. Hence the sign of $
D_C(\eta^R) $ is defined by the sign of $ 3 \,  f_{MF}(\eta^R) - 1 $.

$ D_C(\eta^R) $ is thus {\bf positive} for $ 0 \leq \eta^R < \eta^R_C =
2.517551\ldots $. That is, the mean field for the canonical ensemble
can be applied for $ 0 \leq \eta^R < \eta^R_C $. 

\begin{figure}
\begin{turn}{-90}
\epsfig{file=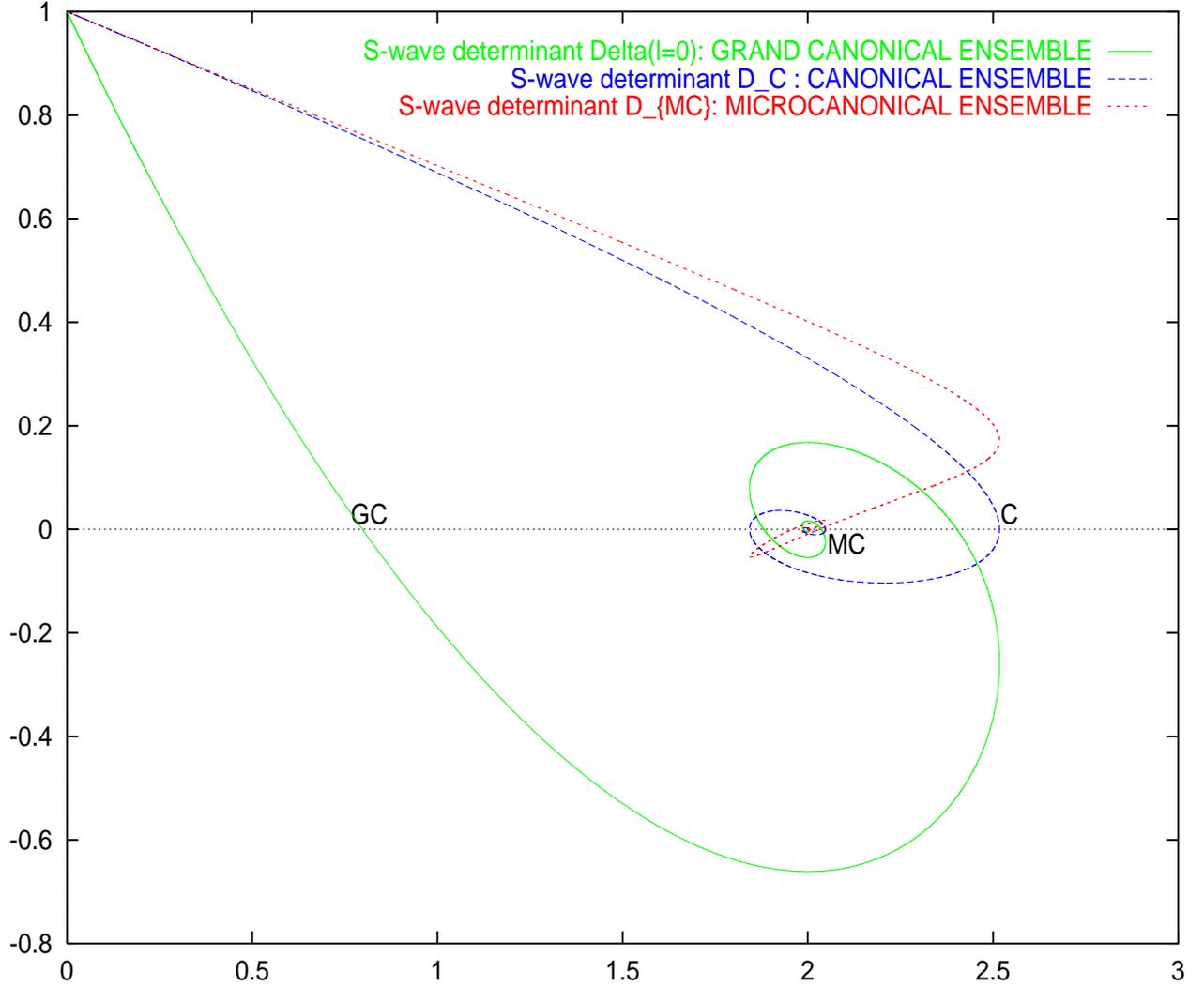,width=14cm,height=18cm} 
\end{turn}
\caption{ 
The S-wave determinants $ \Delta_0(\eta^R), \; D_C(\eta^R) $ and $
D_{MC}(\eta^R) $ in the grand canonical, canonical and
microcanonical ensembles, respectively,  as functions of $ \eta^R
$. Notice that the mean field approximation for  each ensemble breaks
down as soon as the respective determinant becomes negative when $
\eta $ increases starting from $ \eta = 0 $.
\label{fig6}} \end{figure}

$ D_C(\eta^R) $ vanishes linearly in $\sqrt{ \eta^R_C-\eta^R}$ at
$\eta^R=\eta^R_C$. We plot the S-wave part of $ D_C(\eta^R) $ as a
function of $ \eta^R $ in fig. \ref{fig6}. 

In conclusion, the coordinate partition function $ e^{\Phi_N(\eta)} $ 
in the canonical ensemble takes the form
$$
e^{\Phi_N(\eta)}\buildrel{ N>>1}\over= { e^{-N\, s(\eta^R)} \over
\sqrt{  D_C(\eta^R) }}\left[ 1 + {\cal O}\left( {1 \over N}\right)
\right]  
$$

\bigskip

For the various physical quantities, we get  analogous expressions 
to eqs.(\ref{cfluc}), but with Det$_{GC}(\eta^R)$ replaced by $ D_C(\eta^R)
$. That is, in the canonical ensemble, up to the order $ N^{-1} $ the
function $ f(\eta^R) $  takes the form
\begin{equation}\label{fcano}
f(\eta^R) = f_{MF}(\eta^R) + {\eta^R \over 6N}{d \over
d\eta^R}\log D_{C}(\eta^R) +{\cal O}\left({1 \over
N^2}\right) \; .
\end{equation}

The clumping phase transition takes place when $D_C(\eta^R)$ vanishes at
$\eta^R = \eta^R_C$. Near such point the expansion in $1/N$ breaks down
since the correction terms in eq.(\ref{fcano}) become large. 
Mean field applies when $ N|\eta^R_C-\eta^R| >> 1 $.

Since
\begin{equation}\label{fcorcan}
{\eta^R \over 6  }\; { d \over d \eta^R}\log D_C(\eta^R)\buildrel{
\eta^R \uparrow 
\eta^R_C}\over= -{\eta^R_C \over 12(\eta^R_C-\eta^R)} \to -\infty \; ,
\end{equation}
eq.(\ref{fcorcan}) correctly suggests that $PV/[NT], \; E/[NT] $ and
the entropy per particle become {\bf large and negative} for $\eta^R
\uparrow \eta^R_C$. Indeed the Monte Carlo simulations yield a large and
negative value for these three quantities in the collapsed phase
(paper I). 

\subsection{The Microcanonical Ensemble}

We have to compute the Gaussian functional integral in eq.(\ref{gausmic})
\begin{equation}\label{gaumicro}
\int\int DY \, dy_0 \;{ d {\tilde \eta} \over 2 \pi i}\; e^{-N \,
s^{(2)}_{MC}[Y(.),y_0,{\tilde \eta}] } \; 
\end{equation}
where $ s^{(2)}_{MC}[Y(.),y_0,{\tilde \eta}] $ is given by eq.(\ref{s2mc}).

As for the canonical ensemble, we start by finding a saddle point $
{\tilde Y}(.)  $ in the Gaussian functional integral (\ref{gaumicro}). We have,
\begin{equation}\label{micens}
{{\tilde Y}(\vec x) \over \rho({\vec x})} - {\tilde\eta} \int d^3y \;
{{\tilde Y}({\vec y}) + \rho({\vec y}) \over |{\vec x}-{\vec y}|} - y_0 =0 \; ,
\end{equation}
which has as solution
$$
{\tilde Y}(\vec x) = \rho({\vec x}) \left[ \left(  y_0 +
{{\tilde\eta} \over \eta_s} \right) w(\vec x) - {{\tilde\eta} \over
\eta_s} \right] \; .
$$
Here, $ w(\vec x) $ obeys eq.(\ref{ecw}). 

We define a new integration variable $ Z({\vec x}) $ in eq.(\ref{gaumicro}) as
$$
Y(\vec x) = {\tilde Y}(\vec x) + \rho({\vec x}) \; Z({\vec x})
$$
Eq.(\ref{gaumicro}) takes then the form,
\begin{eqnarray}\label{otrop}
\int\int DY \, dy_0 \;{ d {\tilde \eta} \over 2 \pi i}\; e^{-N \,
s^{(2)}_{MC}[Y(.),y_0,{\tilde \eta}] } &=& J \; \int\int DZ \, dy_0
\;{ d {\tilde \eta} \over 2 \pi i}\; e^{-N \left[ A_0 \, (y_0)^2 + B_0
\, {\tilde \eta}^2 + C_0  \, {\tilde \eta}\, y_0 +  A[Z(.)]\right]}
\cr \cr
&=& { \sqrt{3/2} \over \eta \sqrt{\mbox{Det}\left({K_{\rho} \over
K_0}\right)  \,\left[ 4 A_0 \, B_0 - C_0^2 \right]} } \; , \label{resumicro}
\end{eqnarray}
where 
$$
A_0 \equiv -\frac12 \int d^3x \, \rho({\vec x}) \; w({\vec x})
\; , \; B_0 \equiv { 5 \over 4 \eta^2 } -{\xi \over \eta} - \frac12{\eta^2}
\int d^3x \, \rho({\vec x}) \; w({\vec x}) \; , \; C_0 \equiv
\frac1{\eta}\left[ 1 - \int d^3x \, \rho({\vec x}) \; w({\vec
x})\right]
$$
$  A[Z(.)] $ is given by eq.(\ref{adeZ}), $ J $ by  eq.(\ref{jaco})
and we have normalized the integral to unit at $ \eta = 0 $.

Then, the argument of the square-root in eq.(\ref{resumicro}) becomes
\begin{eqnarray}\label{DMCeta}
&&D_{MC}(\eta^R)\equiv { 2 \eta^2 \over 3}\mbox{Det}\left({K_{\rho} \over
K_0}\right) \; \left[ 4 A_0 \, B_0 - C_0^2 \right] \cr \cr
&&= \left[ 6\,  f^2_{MF}(\eta^R) - \left( \frac{11}2 -  \eta^R \right)
f_{MF}(\eta^R) + \frac12 \right] \; 
\prod_{l\geq 1} [\Delta_l(\eta^R)]^{2l+1} \;
e^{-\int d^3x \; \phi(\vec x) } \;  ,
\end{eqnarray}
where we used 
\begin{equation}\label{detsmc}
\frac23 \, (\eta^R)^2 \; \Delta_0(\eta^R)\, \left[
4 A_0 \, B_0 - C_0^2 \right] =  6\,  f^2_{MF}(\eta^R) - \left(
\frac{11}2 -  \eta^R \right) f_{MF}(\eta^R) + \frac12 
\end{equation}
All factors in eq.(\ref{DMCeta}) are positive definite except the
first one. The sign of $ D_{MC}(\eta^R) $ is therefore determined by
the sign of the expression (\ref{detsmc}).

Thus, $ D_{MC}(\eta^R) $ is positive in the interval $ 0 \leq \eta^R \leq
\eta^R_C $ and keeps positive in the second branch of $ f_{MF}(\eta^R)
$ for $ \eta^R_{MC}= 2.03085\ldots < \eta^R \leq \eta^R_C $. At the
point MC ($\eta^R = \eta^R_{MC}= 2.03085\ldots$ in the second Riemann
sheet), the expression (\ref{detsmc}) becomes negative and the mean field
approximation breaks down for the microcanonical ensemble.
$ D_{MC}(\eta^R) $ vanishes linearly in  $ \eta^R - \eta^R_{MC} $ at $
\eta^R = \eta^R_{MC} $. 

\bigskip

In conclusion, the coordinate partition function $ w(\xi,N) $ 
in the microcanonical ensemble takes the form
$$
w(\xi,N)\buildrel{ N>>1}\over= { e^{-N\, s(\eta^R)} \over
\sqrt{  D_{MC}(\eta^R) }}\left[ 1 + {\cal O}\left( {1 \over N}\right)
\right]  
$$

Adding the contributions from the functional determinant to the mean
field results (\ref{abel2}) yields expressions analogous
to eqs.(\ref{cfluc}) but with Det$_{GC}(\eta^R)$ replaced by $ D_{MC}(\eta^R)
$. That is, in the microcanonical ensemble the function $ f(\eta^R) $ to
the order $ N^{-1} $ takes the form,
\begin{equation}\label{fmicro}
f(\eta^R) = f_{MF}(\eta^R) + {\eta^R \over 6N}{d \over
d\eta^R}\log D_{MC}(\eta^R) +{\cal O}\left({1 \over
N^2}\right) \; .
\end{equation}
For $ \eta^R \downarrow\eta^R_{MC} $, reaching the point MC, 
we find
$$
{\eta^R \over 6  }\; { d \over d \eta^R}\log D_C(\eta^R)\buildrel{
\eta^R \downarrow 
\eta^R_{MC}}\over= {\eta^R_{MC} \over 12(\eta^R-\eta^R_{MC})} \to +\infty \; ,
$$
We see that the MF predicts that $ pV/[NT] $ {\bf grows} approaching
the critical point MC. This behaviour is confirmed by the Monte Carlo
simulations. At the point MC, $ pV/[NT] $ increases discontinuously  by
$ 50 \% $ in the Monte Carlo simulations.

\section{Local Physical Magnitudes}

We obtain in this section physical magnitudes at at point $ \vec r $
in the gaseous phase, that is, the space dependence of the particle
density, the local energy density, the pressure and the speed of sound.
We also compute the average distance between particles.

\subsection{Particle Distribution}

The particle distribution at thermal equilibrium obtained through the Monte 
Carlo simulations and mean field methods is inhomogeneous both in the
gaseous and condensed phases.  In the dilute regime $\eta \ll 1$ the gas
density is uniform, as expected. 

We plot in figs. 3-6 in paper I
the density of particles from Monte Carlo simulations in
the cube for the gaseous and for the condensed phases, respectively. 

In the mean field approximation and for the spherically symmetric case,
the particle density is given by
\begin{equation}\label{densr}
\rho_{MF}(r) = e^{\phi(r)} = {\lambda^2 \; e^{\chi(\lambda r)} \over  4\, \pi
\,\eta^R} \quad , \quad 0 \leq r \leq 1\; .
\end{equation}

The mass inside a radius $ r $ is then given by
$$
M(r) = 4\, \pi \int_0^r {r'}^2 \; \rho_{MF}(r')\;dr' = - { \lambda \; r^2
\over\eta^R} \, \chi'(\lambda r) \; ,
$$
where we used eq.(\ref{ecuaxi}). For small $ r $ this gives
\begin{equation}\label{rchico}
M(r)\buildrel{r \ll 1}\over = { \lambda^2 \; r^3 \over3 \, \eta^R}
\left[ 1 + {\cal O}(\lambda^2 \; r^2) \right]\; .
\end{equation}
We find an uniform mass distribution near the origin. This is simply explained
by the absence of gravitational forces at $ r = 0 $. Due to the
spherically symmetry,
the gravitational field exactly vanishes at the origin. The particles
exhibit a perfect gas distribution in the vicinity of $ r = 0
$. Actually, eq.(\ref{rchico}) is {\bf both} a short distance and a
{\bf weak coupling} expression. Eq.(\ref{rchico}) is valid in the
dilute limit $ \eta^R \ll 1 $ for all $ 0 \leq r \leq 1 $.

\begin{figure}[t] 
\begin{turn}{-90}
\epsfig{file=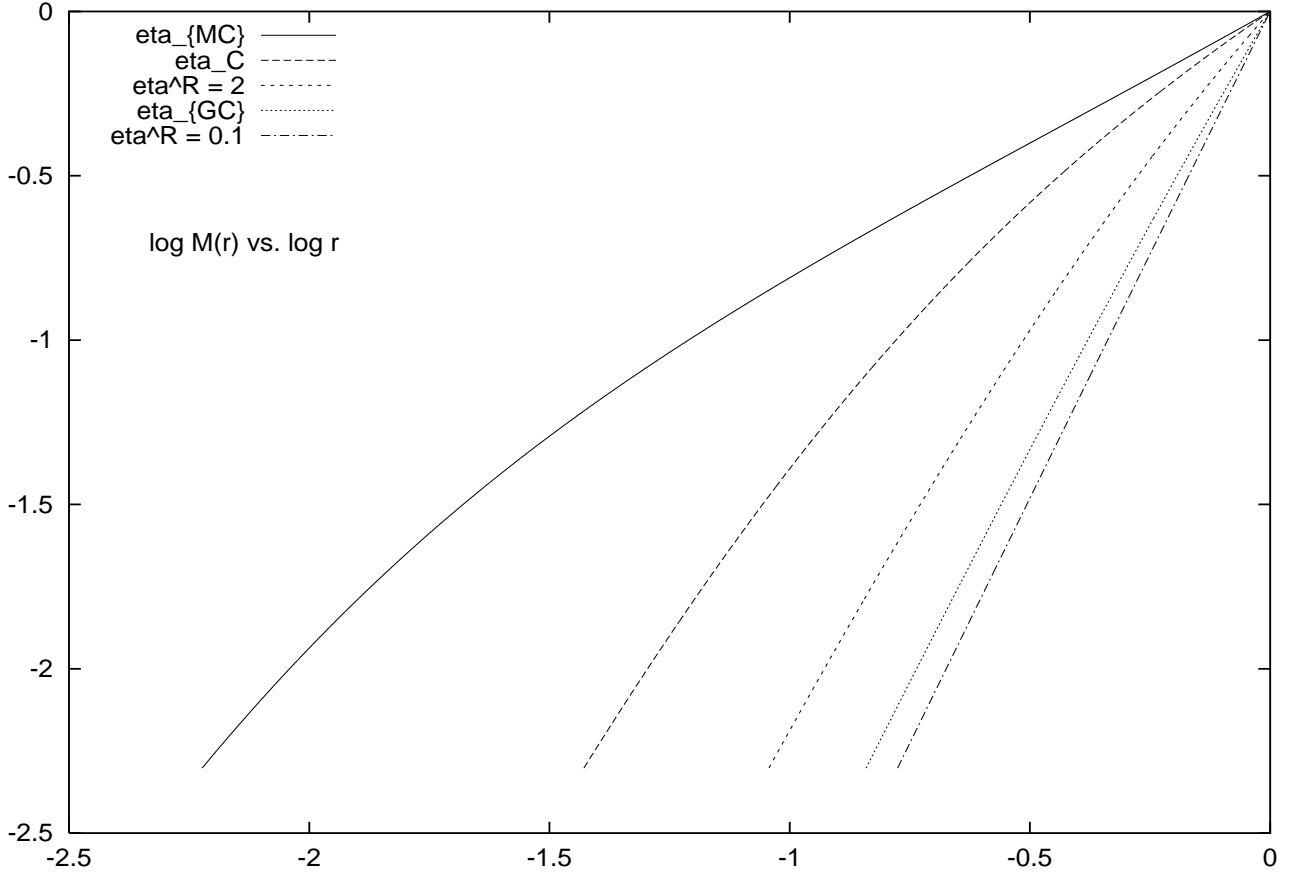,width=12cm,height=18cm} 
\end{turn}
\caption{ $ \log M(r) $ vs. $ \log r $ for five different values of $
\eta^R : \; \eta^R_{MC} = 2.03085\ldots , \; \eta^R_{C} =
2.517551\ldots, \; \eta^R =2 , \; \eta^R_{GC} = 0.797375\ldots $ and
$ \eta^R = 0.1 $. \label{fig12}} 
\end{figure}

\vspace{1cm}

\begin{tabular}{|l|l|l|}\hline
$ \eta^R $ & $\hspace{0.5cm} D $ & $\hspace{0.5cm} C $\\ \hline 
$ 0.1 $ &    \hspace{0.3cm} $ 2.97 $ \hspace{0.3cm}  &  \hspace{0.3cm}
$ 1.0 $ \hspace{0.3cm}  \\ \hline 
$ \eta^R_{GC} = 0.797375\ldots $ & \hspace{0.3cm} $ 2.75  $\hspace{0.3cm} &
\hspace{0.3cm} $ 1.03 $ \hspace{0.3cm} \\ \hline 
 $ 2.0 $  & \hspace{0.3cm} $ 2.22 $ \hspace{0.3cm} &\hspace{0.3cm} $ 1.1
$\hspace{0.3cm} \\ \hline 
$ \eta^R_{C} = 2.517551\ldots $ & \hspace{0.3cm} $ 1.60
$\hspace{0.3cm} & \hspace{0.3cm} 
$ 1.07 \hspace{0.3cm} $ \\ \hline
$ \eta^R_{MC} = 2.03085\ldots $ & \hspace{0.3cm} $ 0.98  $
\hspace{0.3cm}&\hspace{0.3cm} 
$ 1.11 $\hspace{0.3cm} \\ \hline 
\end{tabular}

\vspace{0.5cm}

{TABLE 2. The Fractal Dimension $ D $ and the proportionality
coefficient $ C $ as a function of $ \eta^R $ from a fit to the mean
field results according to $ { \cal M }(r) \simeq C \; r^D $.}

\vspace{0.5cm}

We plot in fig. \ref{fig12} the particle distribution for $ 90 \% $ of
the particles for several values of $ \eta^R $. We exclude in the
plots the region $ M(r) < 0.1 $ where the distribution is uniform.

We find that these mass distributions approximately follow the power law
\begin{equation}\label{esca}
{ \cal M }(r) \simeq C \; r^D 
\end{equation}
where, as depicted in Table 2, $ D $ slowly decreases with  $
\lambda(\eta^R) $ from the value $ D = 3 $ for the ideal gas ($\eta=0$) till $
D = 0.98 $ in the extreme limit of the MC point.

The presence of a critical region where scaling holds supports our
previous work in the grand canonical ensemble  based on field
theory\cite{natu,prd,gal,eri}. 

\bigskip

The values of the fractal dimension $ D $ of the self-gravitating gas
are around $ D= 2 $ (see Table 2). Such value 
can be analytically obtained assuming exact scale invariance,
the virial theorem and the extensivity of the total energy in the
limit defined by eq.(\ref{limiT}) as follows, \cite{claudio}.

Let us assume that the density scales as $ \rho(r) \sim r^{-a}
$. Then, using the virial theorem, the total energy $ E $ will scale as 
$$
E \sim \int^R d^3r\;  d^3r'\;{  \rho(r)\; \rho(r') \over |{\vec r} -{\vec
r'}|} \sim R^{5-2a} 
$$
Therefore,
$$
{ E \over R^3} \sim R^{2(1-a)}
$$
Now, extensivity requires $ E/V $ to be independent of $ R
$, that is, $ a = 1 $ and 
$$
\rho(r)\sim r^{-1} \quad  , \quad M(R) \sim R^2 \; .
$$

In addition, this implies that the gravity force in the surface of the
gas is independent of $ R $.

\subsection{Average distance between particles}

We investigate here the average distance between particles $ <r> $ and
the average squared distance $ <r^2> $. The study of $ <r> $ and $
<r^2> $ as functions of $ \eta $ permit a better understanding of the
self-gravitating gas and its phase transition in the different
statistical ensembles. 

These average distances are defined as
\begin{eqnarray}\label{defryr2}
<r> &\equiv& \int \int\left| {\vec r}- {\vec r \,}' \right| \; <
\rho({\vec r}) \; 
\rho({\vec r \, }') > \; d^3 r \; d^3 r'  \; , \cr \cr
<r^2> &\equiv& \int \int\left| {\vec r}- {\vec r \, }' \right|^2  \; <
\rho({\vec r}) \; \rho({\vec r \, }') > \; d^3 r \; d^3 r'
\end{eqnarray}
In the mean field approximation we have 
\begin{equation}\label{correMF}
< \rho({\vec r}) \; \rho({\vec r \, }') > = \rho_{MF}({\vec r}) \;
\rho_{MF}({\vec r \, }') + {\cal O}\left( {1 \over N} \right) \; .
\end{equation}
In addition, in the spherically symmetric case we use eq.(\ref{densr})
for the particle density. In Appendix E we compute the 
integrals in eqs.(\ref{defryr2}) and we get as result,
\begin{eqnarray}\label{ryr2exp}
<r> &=& 2 - 2 \; \int_0^1 r^2 \; dr \; \left[ 1 + { \phi(r) -\phi(1)
\over \eta^R} \right]^2  \; , \cr \cr
<r^2> &=& 2 - {12 \over \eta^R}\; \int_0^1 r^2 \; dr \; \left[ 
\phi(r) -\phi(1) \right]
\end{eqnarray}
where $\phi(r)$ is given by eq.(\ref{fixi}).

\begin{figure}[t] 
\begin{turn}{-90}
\epsfig{file=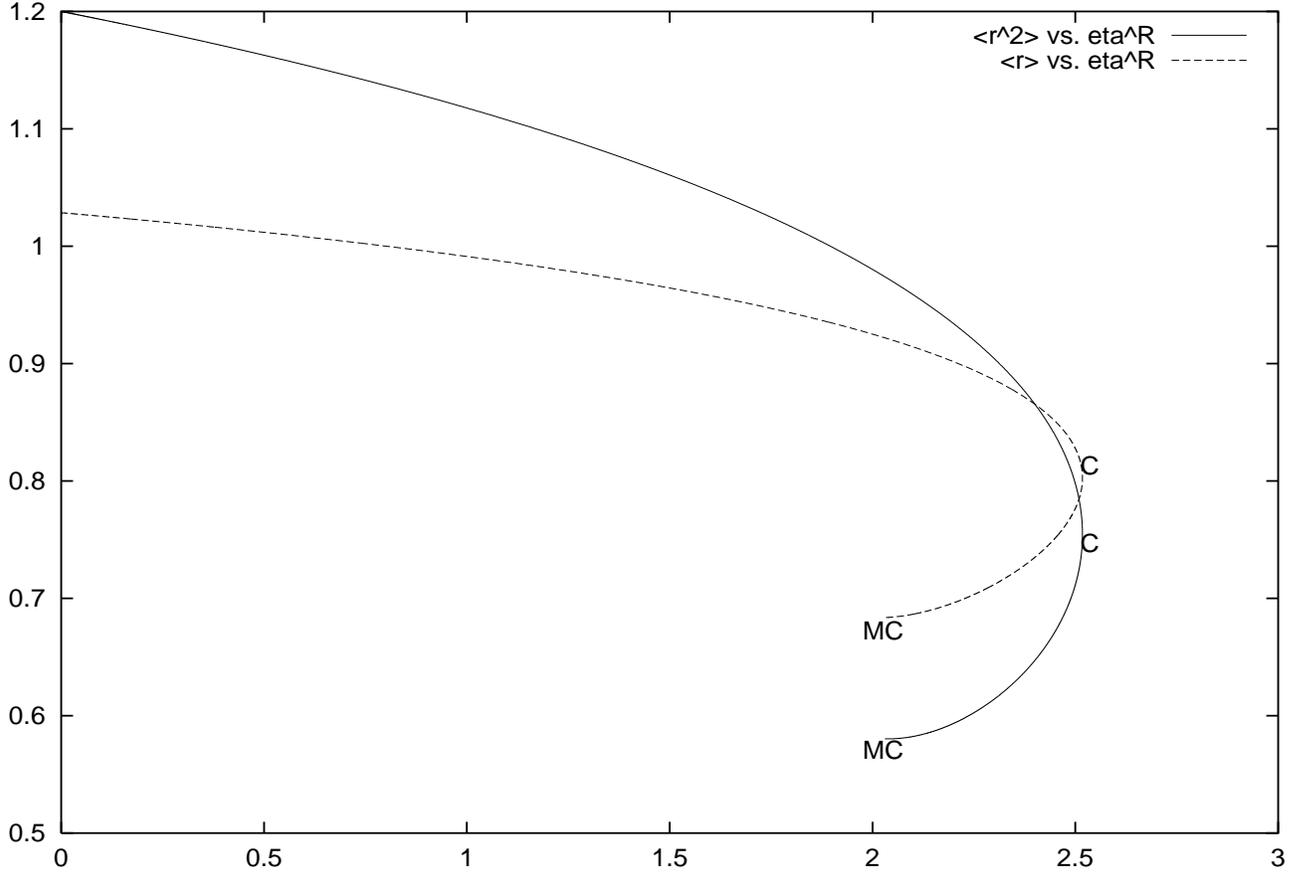,width=12cm,height=18cm} 
\end{turn}
\caption{Mean value of the distance between particles ($ <r> $) and
mean value of the squared distance between particles
($ <r^2> $) as functions of $ \eta^R $ in the mean
field approach from eqs.(\ref{ryr2exp}). Notice that the particles are
inside a sphere of unit radius. 
\label{fig15}} 
\end{figure}

We plot $ \, <r> $ and $ <r^2> $ as functions of $
\eta^R $ in fig. \ref{fig15}. Both $ <r> $ and $ <r^2> $
monotonically decrease with $ \lambda(\eta^R) $. Their values for the
ideal gas are 
$$
\left. <r>\right|_{\eta=0} = \frac{36}{35}=1.02857\ldots \quad , \quad\left.
<r^2>\right|_{\eta=0} = \frac65 \; .
$$
At the critical points (C for the canonical ensemble and MC for the
microcanonical ensemble) the average distances
sharply decrease. Both  $ \, <r> $ and $ <r^2> $ have infinite
slope as functions of $\eta^R $ at the point C.

\begin{figure}[t] 
\begin{turn}{-90}
\epsfig{file=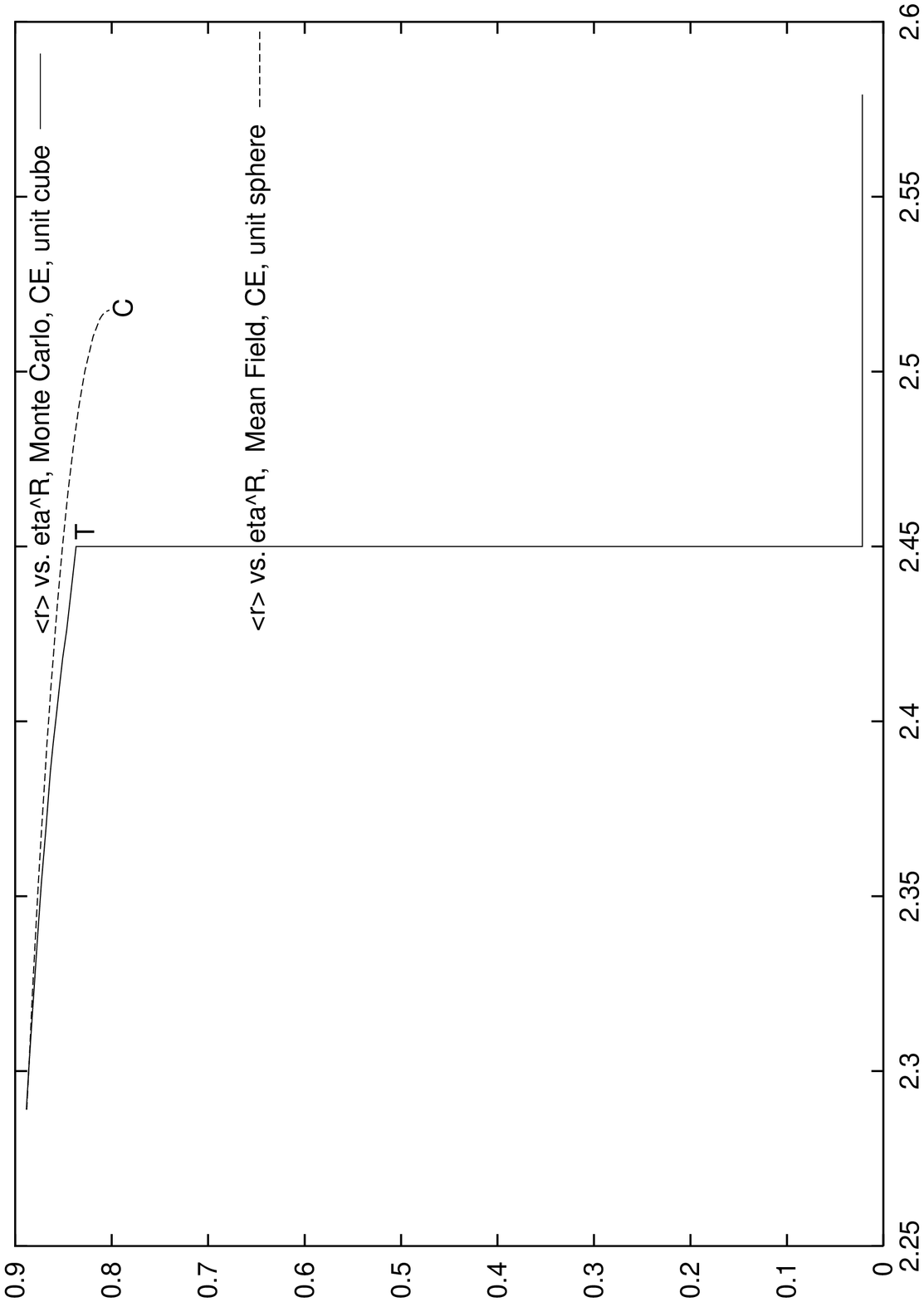,width=12cm,height=18cm} 
\end{turn}
\caption{Mean value of the distance between particles ($ <r> $) in a unit cube
from Monte Carlo simulations and in a unit sphere from mean field as
functions of $ \eta^R $. 
\label{rmmcmf}} 
\end{figure}

We plot in fig. \ref{rmmcmf} the Monte Carlo results for $ <r> $ in a unit cube
together with the MF results in a unit sphere. Notice that $ <r> $
sharply falls at the point T clearly indicating the transition to collapse. 

\subsection{Local energy density and gravitational potential}

The gravitational potential at the point $ \vec q $ is given by
\begin{equation}\label{udeq}
U({\vec q }) = - G \, m \sum_{1\leq l \leq N} 
{1 \over { |{\vec q} - {\vec q}_l|}} = - {G \, m \over L } \sum_{1\leq
l \leq N}  {1 \over { |{\vec r} - {\vec r}_l|}}
\end{equation}
where $ {\vec q} = L {\vec r} $ and  $ {\vec q}_l = L {\vec r}_l ,
1\leq l \leq N . \;  U({\vec q }) $ can be easily related to the
saddle point solution in the mean field approach. We write the sum in
eq.(\ref{udeq}) in terms of the particle density $ \rho_s({\vec r}) =
e^{\phi({\vec r})} $ as,
\begin{equation}\label{uder}
U({\vec q }) =  - {G \, m \, N \over L }
\int  { d^3y \over |{\vec r}-{\vec y}|} \;  \rho({\vec y})
\end{equation}
Comparison of the saddle point equation (VI.24) in paper I and
(\ref{uder}) yields, 
\begin{equation}\label{ufi}
U({\vec q }) =  - { T \over m} \left[ \phi({\vec r}) - a_s \right] 
\end{equation}
and using eq.(VI.67)  in paper I we recover the relation \cite{prd}
$$
U({\vec q }) =  - { T \over m}\; \Phi_s({\vec r}) \; .
$$

\bigskip

The local density of potential energy is thus given by,
$$
\epsilon_P({\vec r}) = \frac12 \; m \; \rho({\vec q}) \;U({\vec q })=
- {NT \over 2V} \; \left[ \phi({\vec r}) - a_s \right] \;
e^{\phi({\vec r})} \; ,
$$
while the local density of kinetic energy takes the form
$$
\epsilon_K({\vec r}) = {3NT \over 2V} \;e^{\phi({\vec r})} \; .
$$
It is easy to check that
$$
U = \int d^3 q \; \epsilon_P({\vec r}) \quad , \quad \frac32 N T = \int
d^3 q  \; \epsilon_K({\vec r}) \; .
$$
where $ U = E - 3NT/2 $ follows from eq.(\ref{abel2}).

In the spherically symmetric case, the local energy density takes the form
\begin{equation}
\epsilon(r) = \epsilon_K(r) + \epsilon_P(r) = {NT \over V} \; {
\lambda^2 \over 8 \, \pi \, \eta^R } \left[ 3 - \eta^R + \chi(\lambda) -
\chi(\lambda\,r) \right]\; e^{\chi(\lambda\,r) } \; ,
\end{equation}
where we used eqs.(\ref{fixi}), (\ref{fide1}) and eq.(VI.55) from paper I.

The energy density at the surface is always positive:
$$
\epsilon(1) ={NT \over V} \; {\lambda^2 \over 8 \, \pi \, \eta^R } \; (3
- \eta^R ) > 0\; ,
$$
whereas the energy density at the center,
$$
\epsilon(0) = {NT \over V} \; {\lambda^2 \over 8 \, \pi \, \eta^R }
\left[ 3 - \eta^R + \chi(\lambda) \right]\; ,
$$
is positive for $ 0 \leq \eta^R <  \eta_3^R = 1.73745\ldots $ and
negative beyond the point $ \eta^R = \eta_3^R = 1.73745\ldots $.

\begin{figure}
\begin{turn}{-90}
\epsfig{file=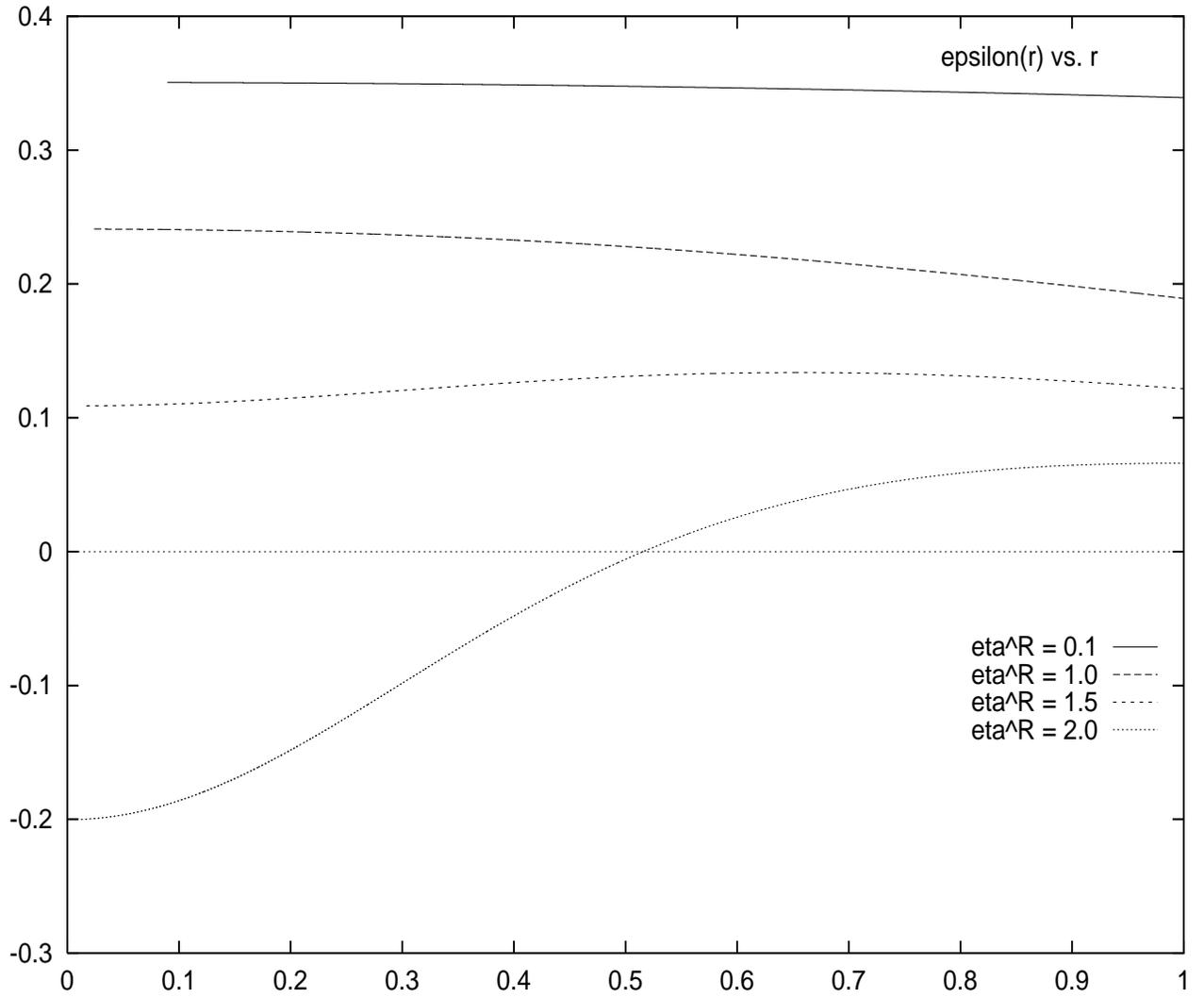,width=14cm,height=18cm} 
\end{turn}
\caption{ The local energy density, $ \epsilon(r) $ as a function of $
r $ in units of $ {NT 
\over V} $ for $ \eta^R= 0.1, \; 1.0, \; 1.5 $ and $ 2.0 $. 
\label{eden1}}
\end{figure}

\begin{figure}
\begin{turn}{-90}
\epsfig{file=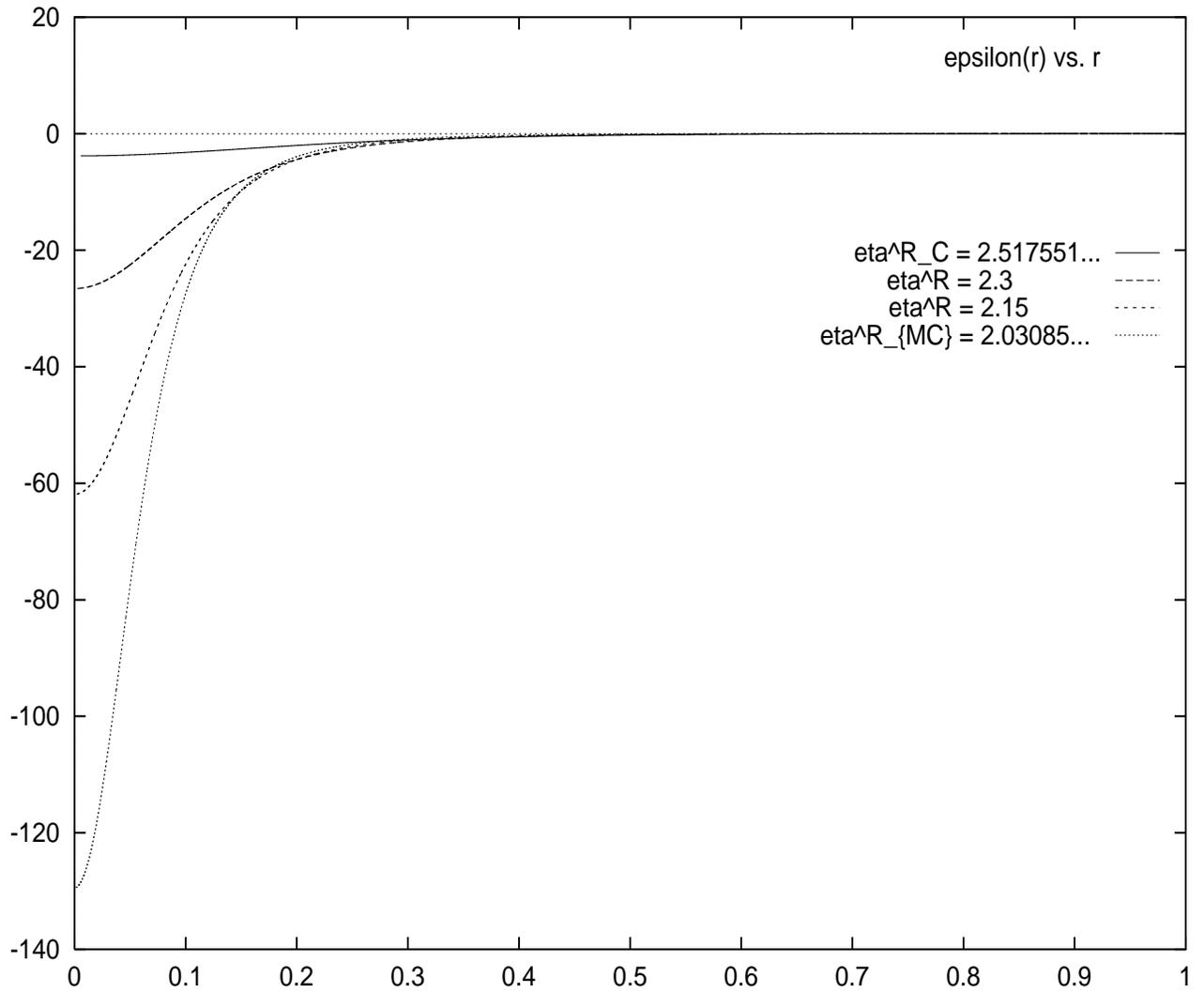,width=14cm,height=18cm} 
\end{turn}
\caption{ The local energy density, $ \epsilon(r) $ as a function of $
r $ in units of $ {NT 
\over V} $ for $ \eta^R $ in the second sheet $ \eta^R= \eta^R_C, \;
2.3, \; 2.15 $ and $ \eta^R_{MC} $. 
\label{eden}}
\end{figure}

We plot in figs. \ref{eden1} and \ref{eden} the energy density as a
function of $ r $ for different values of $ \eta^R $ in the first and
second sheets. 

\subsection{The pressure at a point ${\vec r}$ and the local equation
of state}

The pressure $ p $ that we have obtained in sec. VII [see
fig. 9 in paper I] corresponds to the external pressure on the gas. Let
us now calculate the local pressure at a point $ {\vec r} $ in the interior
of the self-gravitating gas. We perform that computation in the mean field
approach. 

The gravitational force is given by 
\begin{equation}\label{fuerza}
F({\vec q}) = -m \; \rho({\vec q}) \; {\vec \nabla}_q U({\vec q}) = {T
\, N \over L^4} \; e^{\phi({\vec r})} \;  {\vec \nabla}_r\phi({\vec
r}) =  {T \, N \over L^4} \; {\vec \nabla}_r  e^{\phi({\vec r})} \; ,
\end{equation}
where we used eq.(\ref{ufi}), the expression for the particle density
in the volume $ L^3  , \; \rho_V({\vec q}) = {N \over L^3 } \;
e^{\phi({\vec r})} $ and $ {\vec q} = L \; {\vec r} $. 

Since the density of force eq.(\ref{fuerza}) is the gradient of the local
pressure, we find
\begin{equation}
 {\vec \nabla}_r p({\vec r})  = {T \, N \over L^3} \; {\vec
\nabla}_r e^{\phi({\vec r})} = T \;  {\vec \nabla}_r \rho_V({\vec r})
\end{equation}
i. e.
\begin{equation}\label{pr}
p({\vec r}) = T \; \rho_V({\vec r})
\end{equation}
That is, we have shown that the equation of state for the
self-gravitating gas is {\bf locally} the {\bf ideal gas } equation in the mean
field approximation. Notice that contrary to ideal gases, the density
here is {\bf never uniform} in thermal equilibrium. Therefore, in
general the pressure at the surface of a given volume is not equal
to the temperature times the average density of particles in the
volume. In particular, for the whole volume, $ PV/[NT] < 1 $ (except for $
\eta = 0 $).

The local pressure in the spherically symmetric case can be written in
a more explicit way using eqs.(\ref{fixi}) and (\ref{pr}):
\begin{equation}\label{prad}
{ p(r) \; V \over N \, T} = {\lambda^2 \over 3 \, \eta^R}\;
e^{\chi(\lambda\, r)}
\end{equation}
For $ r=1 $, eqs. (\ref{fetexpl}) and (\ref{abel2}) show that $ p = p(1) $
coincides with the external pressure.

The local density and the local pressure monotonically decreases with $
r $. 

\begin{figure}
\begin{turn}{-90}
\epsfig{file=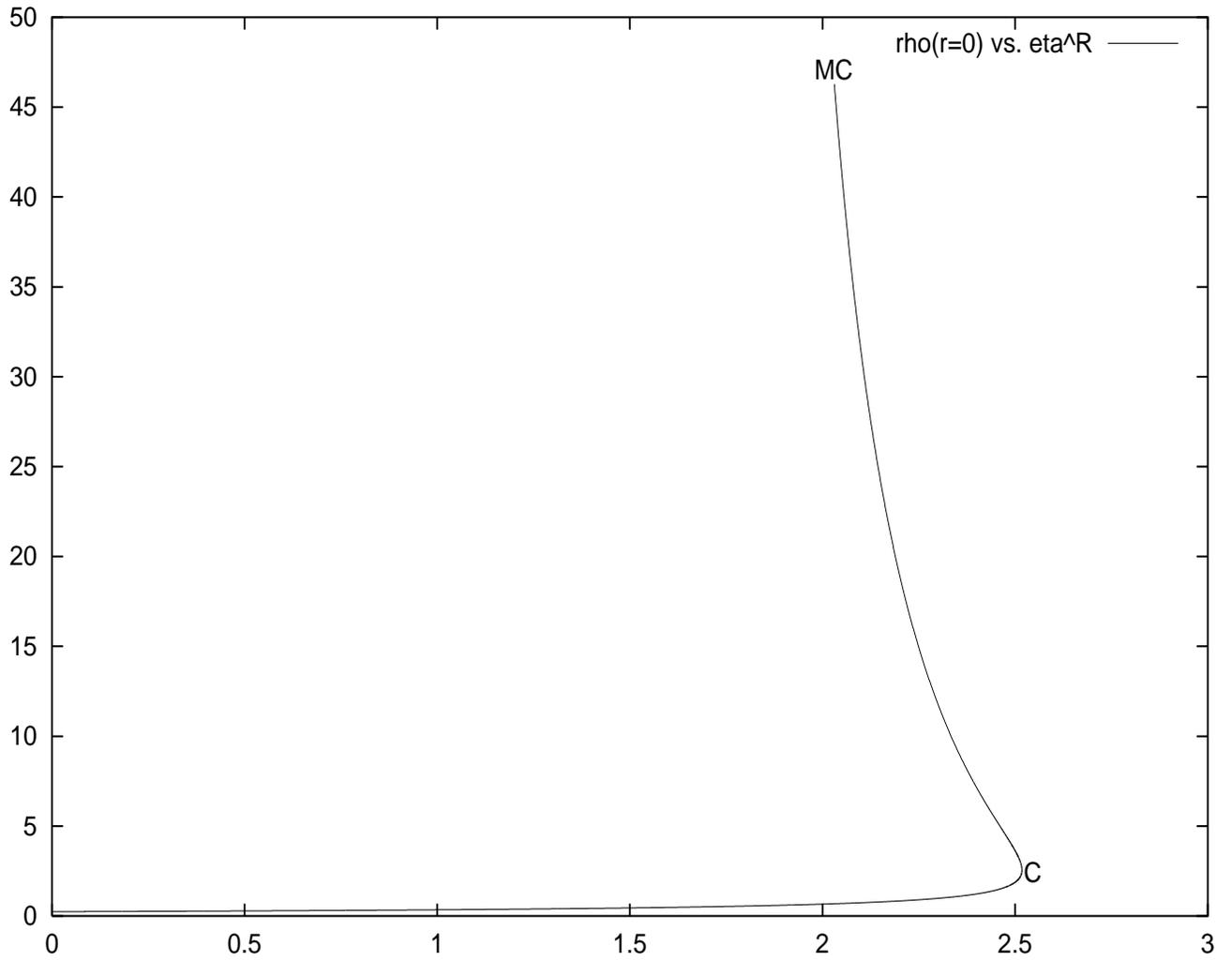,width=14cm,height=18cm} 
\end{turn}
\caption{ $\rho(r=0) = { \lambda^2 \over 4 \, \pi \; \eta^R}$ as a
function of $ \eta^R $. For $ \eta^R = 0 $, 
$\rho(r=0) = 3 /(4 \pi) $ [see eq.(\ref{rhoid})].
\label{rho0}}
\end{figure}

The particle density at the origin follows from eqs.(\ref{densr}) and
(\ref{ecuaxi}):
\begin{equation}\label{rhorig}
\rho(r=0) = { \lambda^2 \over 4 \, \pi \; \eta^R}
\end{equation}
This particle density at the origin
{\bf grows} when moving from $ \eta = 0 $ to MC as shown in
fig. \ref{rho0}. In particular, for $ \eta^R = 0 $ we have
\begin{equation}\label{rhoid}
\left. \rho(r=0)\right|_{\eta^R = 0} = { 3 \over 4 \pi }
\end{equation}
where we used eqs. (\ref{fetexpl}), (\ref{eta0}) and (\ref{rhorig}).
Notice that $ \rho(r) $ is $r$-independent for $ \eta^R = 0 $.

The particle density at the 
surface is proportional to $ f_{MF}(\eta^R) $ [see eq.(\ref{fide1})]
and plotted in fig. 9 of paper I. We see that it {\bf decreases} when moving
from $ \eta = 0 $ to $ \eta = \eta_{Min} = 2.20731\ldots $ in the
second sheet. The migration of particles towards the center as $ \eta
$ varies is manifestly responsible for these variations in the density.

\begin{figure}
\begin{turn}{-90}
\epsfig{file=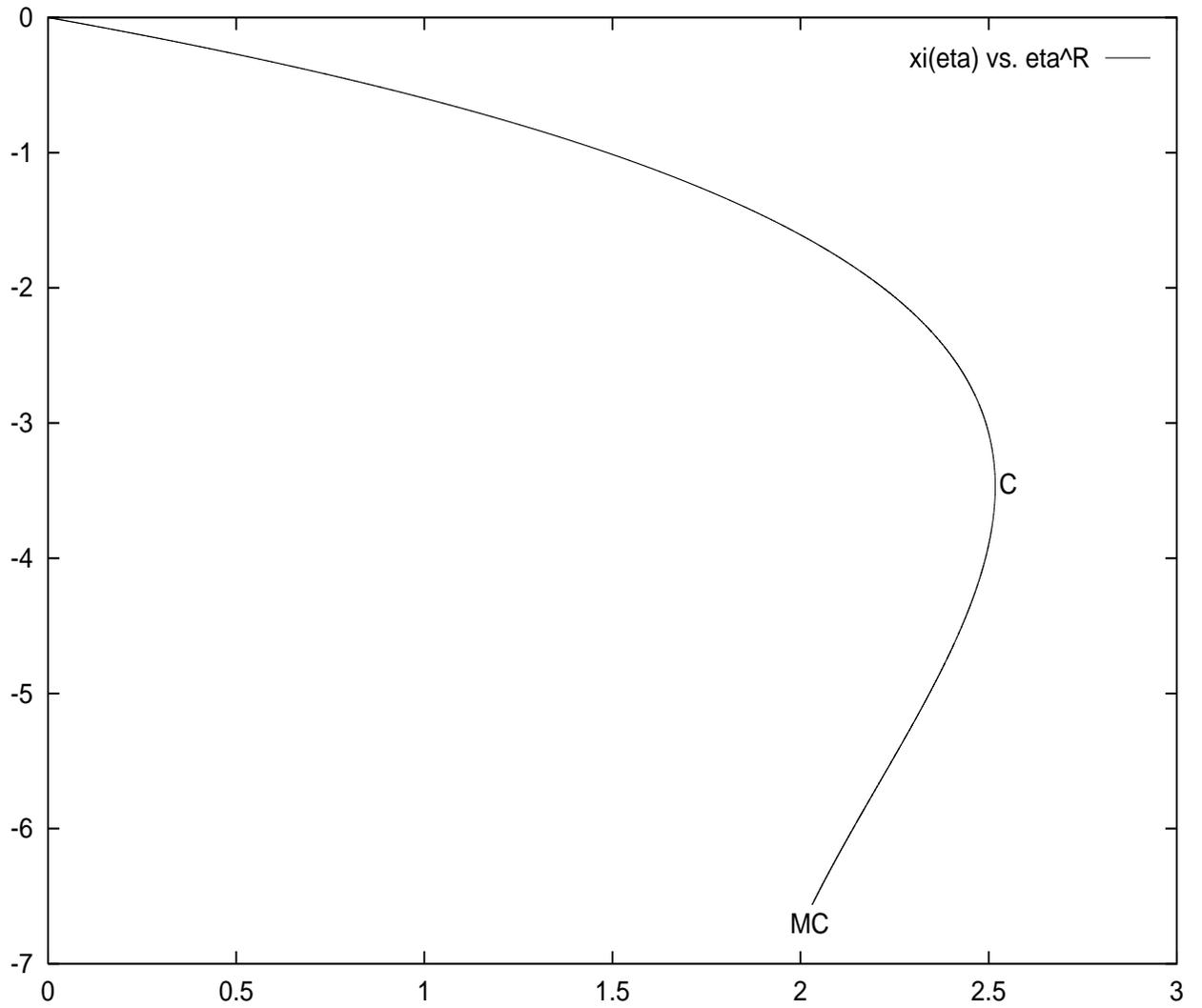,width=14cm,height=18cm} 
\end{turn}
\caption{ $\chi(\lambda(\eta^R)) = \log { p(0) \over p(1) } = \log {
\rho(0) \over \rho(1) } $ as a function of $ \eta^R $. 
\label{fxieta}}
\end{figure}

The pressure (and density) contrast is given by
$$
{ p(0) \over p(1) } = { \rho(0) \over \rho(1) } = e^{-\chi(\lambda)}
$$
We plot in fig. \ref{fxieta} $ \chi(\lambda(\eta^R)) $ as a function of
$ \eta^R $. For $ \eta^R = \eta^R_{C} $ and  $ \eta^R = \eta^R_{MC} $
we recover the known values $ p(0) / p(1) = 32.125\ldots  $ and $ p(0) / p(1) =
708.63\ldots $, respectively \cite{pad,HK}.

\bigskip

Notice that $ p V/[NT] < 1 $ (see fig.  9 in paper I) for $ \eta > 0 $ although
the equation of state is {\bf locally} the one of an ideal gas as we
have showed. The inhomogeneous particle distribution in the
self-gravitating gas is responsible of such inequality. 

Local equations of state other than the ideal gas are often assumed in
the context of self-gravitating fluids
\cite{chandra,sas,lynbell,pad}. Our result imply that forces other
than gravitational are necessary to obtain a non-ideal local equation
of state in thermal equilibrium. 

We have thus shown the equivalence for the self-gravitating gas between
the statistical mechanical treatment in the mean 
field approach with the hydrostatic description \cite{sas}-\cite{bt}.

\subsection{The speed of sound as a function of  ${\vec r}$}

For very short
wavelengths $ \lambda_s \ll L $, the sound waves just feel the local
equation of state (\ref{pr}) and the speed of sound will be that of an
ideal gas. For long wavelengths (of the order $L$), the situation
changes. The calculation in eq.(\ref{vsmf}) corresponds to the speed of sound
for an external wave arriving on the sphere in the long wavelength limit. 
Let us now make the analogous calculation for a wave reaching the
point $ {\vec q} $ inside the gas.

Our starting point is again eq.(III.29) in paper I, 
\begin{equation}\label{vsr}
v_s^2({\vec q}) = - {{ c_P \; V^2} \over {c_V \; N}} \left({ \partial
p({\vec q}) \over \partial V}\right)_{T,{\vec q}} \; ,
\end{equation}
where $ c_P $ and $ c_V $ are the specific heats of the whole system
at constant (external) pressure and volume, respectively, and $
p({\vec q}) $ is the local pressure at the point $ {\vec q} $.

We find for the spherically symmetrical case in MF
\begin{equation}\label{vsre}
{v_s^2(r) \over T} = { c_P \over c_V} \; { \lambda^2 \over 9 \, \eta^R
\; \left[3 f(\eta^R) -1\right] } \; \left[ 6 \, f(\eta^R) + \lambda \, r \;
\chi'(\lambda\, r)\right] \; e^{\chi(\lambda\, r)}\; ,
\end{equation}
where we used eqs.(\ref{prad}), (\ref{vsr}) and
$$
\left({\partial \eta \over \partial V}\right)_T = - {\eta \over 3 \, V} \; .
$$
[$\lambda$ is a function of $\eta^R$ as defined by eq.(\ref{lambaxi})].

At the surface, $ (r=1) , \;  v_s^2(r) $ reduces to eq.(\ref{vsmf}) after using
eqs.(\ref{lambaxi}), (\ref{fetexpl}), (\ref{cvmf}) and (\ref{cpmf}).

\begin{figure}
\begin{turn}{-90}
\epsfig{file=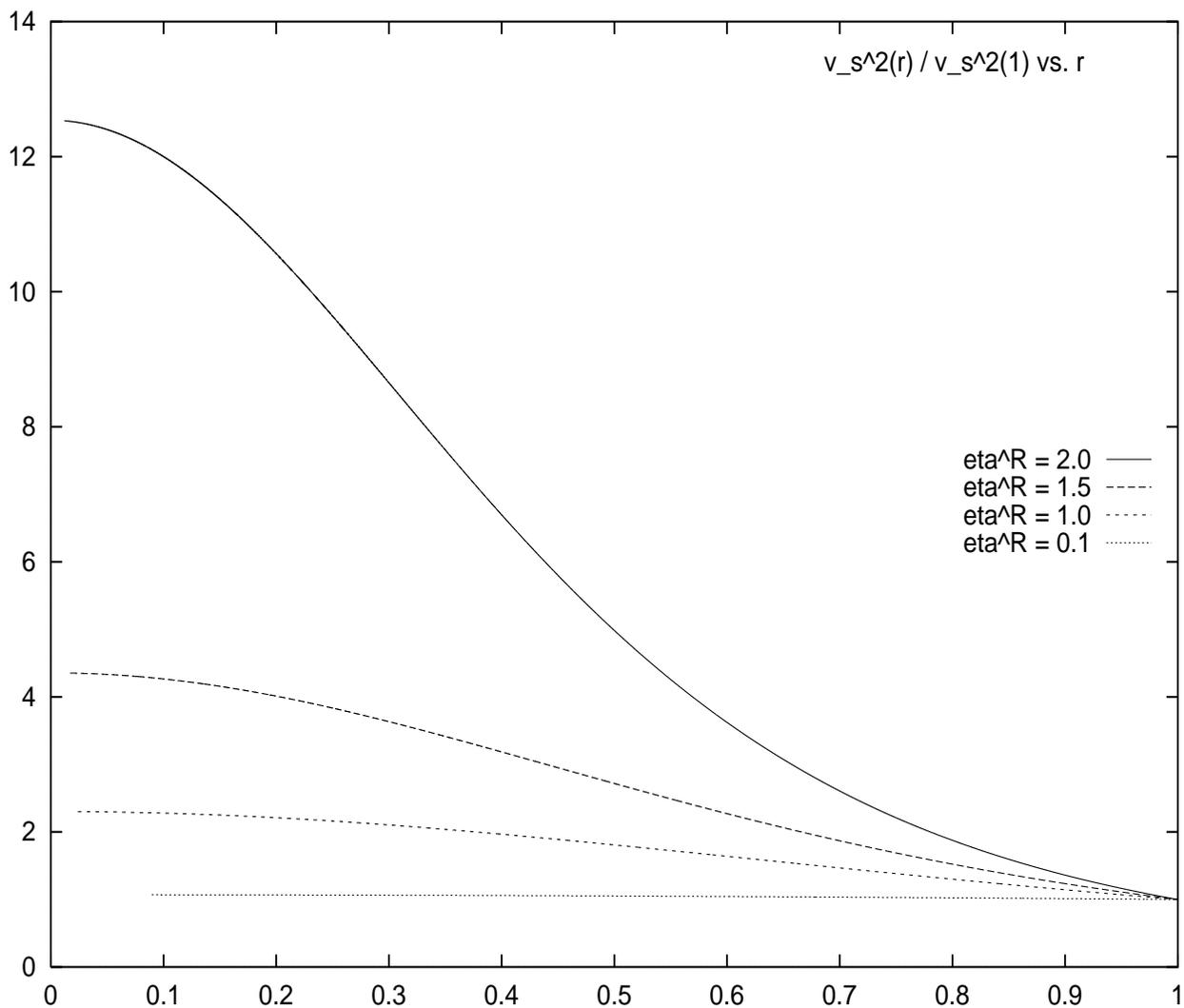,width=14cm,height=18cm} 
\end{turn}
\caption{ $ v_s^2(r) / v_s^2(1)  $ as a function of $ r $  for
$\eta^R= 2.0, \; 1.5, \; 1.0 $ and $ 0.1 $. That is, 
values of $\eta^R$ smaller than  $\eta^R_0=2.43450\ldots$. $ v_s^2(r) /
v_s^2(1)  $ is here always positive and decreases with $ r $.
\label{vsonrinf}}
\end{figure}

\begin{figure}
\begin{turn}{-90}
\epsfig{file=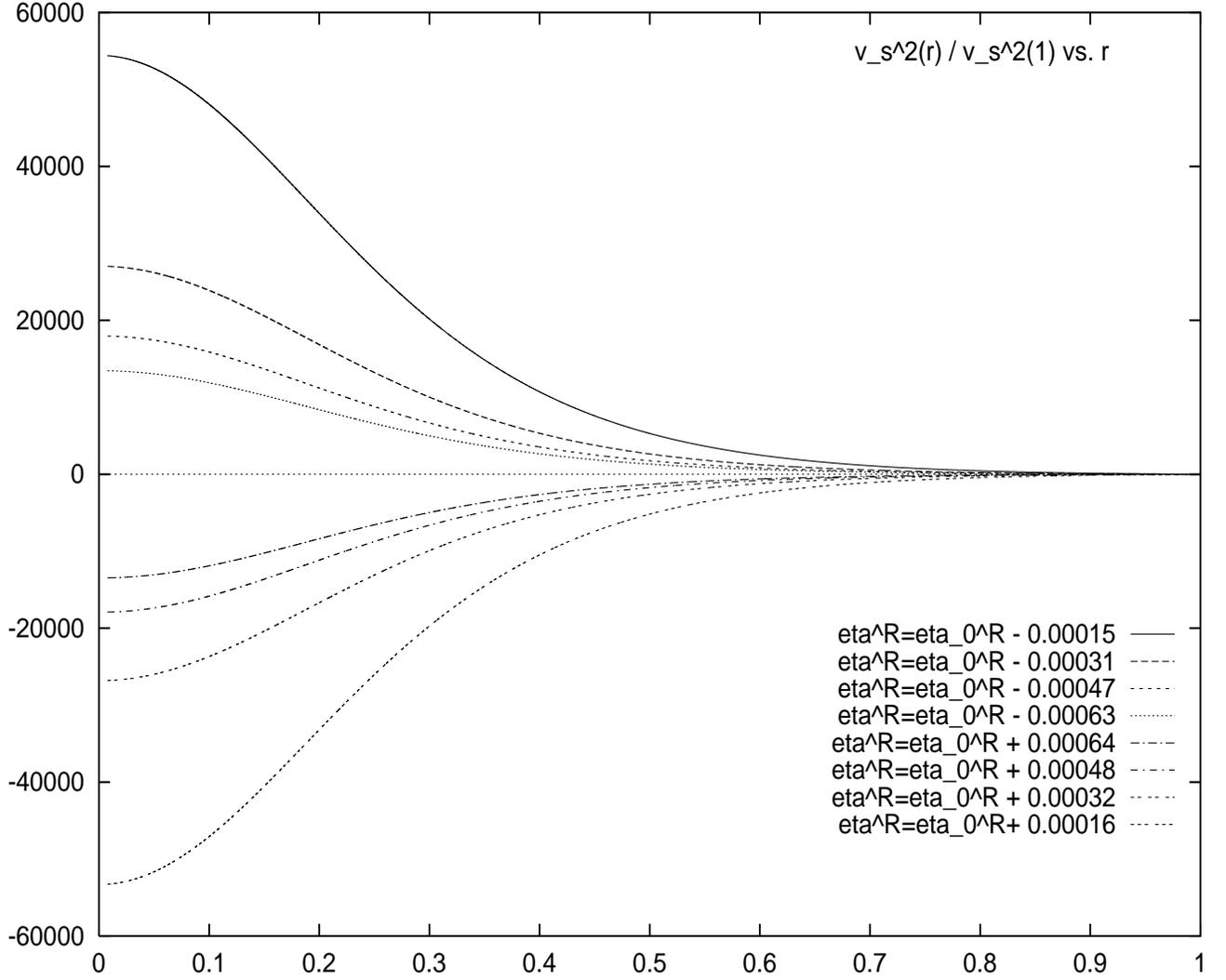,width=14cm,height=18cm} 
\end{turn}
\caption{ $ v_s^2(r) / v_s^2(1)  $ as a function of $ r $  for
values of $\eta^R$ around $\eta^R_0=2.43450\ldots$. Positive values of
$ v_s^2(r)/ v_s^2(1)  $ correspond to $\eta^R < \eta^R_0$ and negative
values of $ v_s^2(r)/ v_s^2(1)  $ correspond to $\eta^R >
\eta^R_0$. We see that the speed of sound squared tends to $ +\infty $
%in the bulk ($ r \lesssim 1 $) for $ \eta^R \uparrow \eta^R_0 $ while it 
in the bulk ($ r < 1 $) for $ \eta^R \uparrow \eta^R_0 $ while it 
tends to $ -\infty $ for $ \eta^R \downarrow \eta^R_0 $.
\label{vsonr}}
\end{figure}

\begin{figure}
\begin{turn}{-90}
\epsfig{file=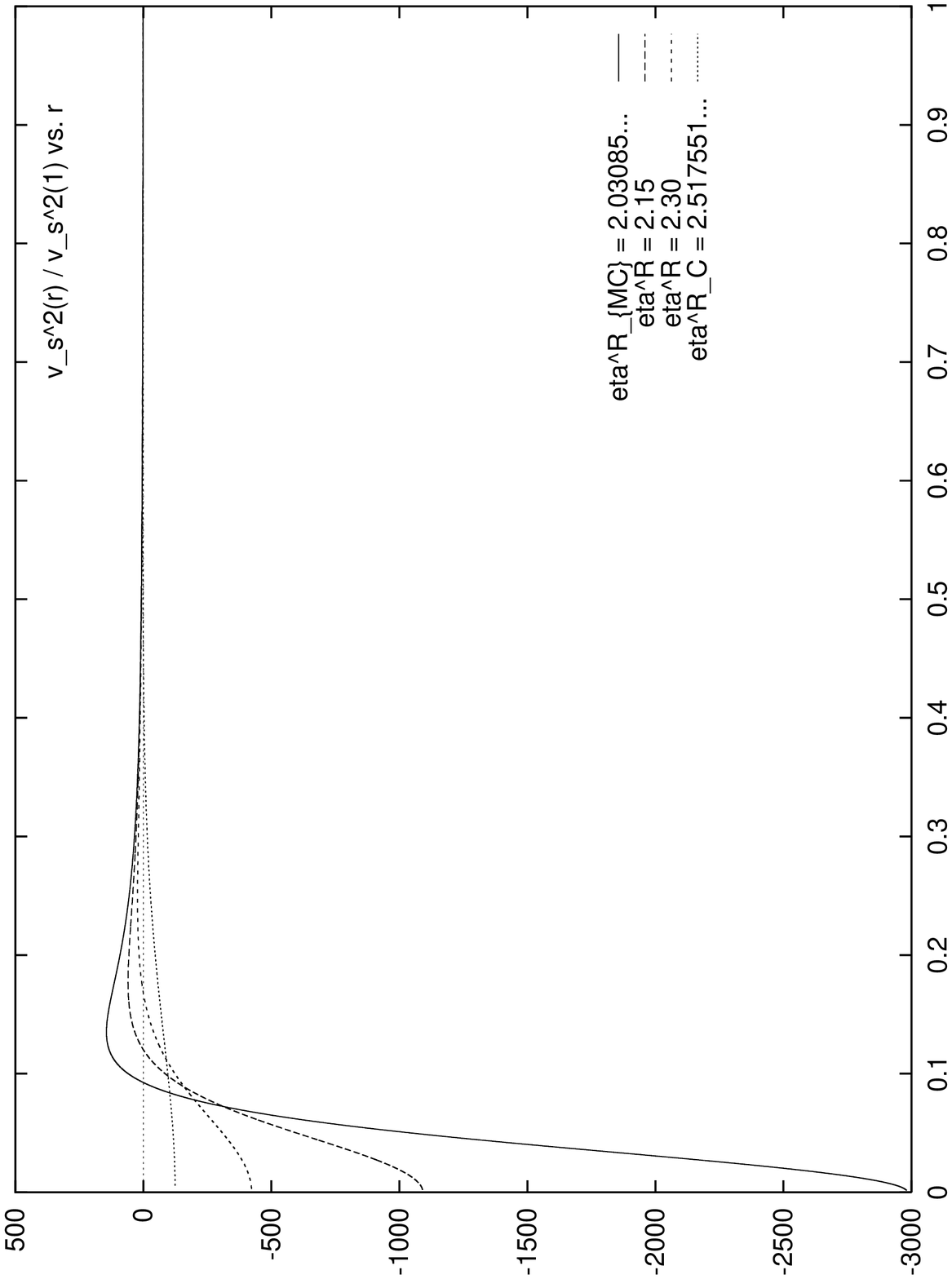,width=14cm,height=18cm} 
\end{turn}
\caption{ $ v_s^2(r) / v_s^2(1)  $ as a function of $ r $  for
 $\eta^R$ in the second sheet: 
$\eta^R= \eta^R_{MC} = 2.03085\ldots, \; 2.15, \; 2.3 $ and $
\eta^R_C = 2.517551\ldots $. $ v_s^2(r) / v_s^2(1) $ is here strongly
negative in the core of the 
sphere. Notice that $ v_s^2(1) < 0 $ for  $ \eta^R = \eta^R_{MC} =
2.03085\ldots $ while  $ v_s^2(1) > 0 $ for  $ \eta^R = 2.15, \; 2.3 $
and $ \eta^R_C $. 
\label{vsonrsup}}
\end{figure}

For $ \eta^R < \eta^R_0 = 2.43450\ldots $ in the first sheet, $
v_s^2(r) $ is positive and decreases with $ r $ as shown in
fig. \ref{vsonrinf}.  

At $ \eta^R
= \eta^R_0 , \; v_s^2(r) $ diverges for all $ 0 \leq r <1 $ due to the factor $
c_P $ in eq.(\ref{vsre}) [cfr. eq.(\ref{cpmf})]. The
derivative of $p$ with respect to $V$ is proportional to $ 6 \,
f(\eta^R) + \lambda \, r \; \chi'(\lambda\, r) $ as we see in
eq.(\ref{vsre}). At $ r=1 $ this 
factor becomes $ 6 \, f(\eta^R) - \eta^R $ [see eq.(\ref{lambaxi})]
which exactly vanishes at $ \eta^R = \eta^R_0 $ canceling the
singularity that $ c_P $ possesses at such point [see eq.(\ref{cpmf})].  
Thus, $ v_s^2(1) $ is regular at $ \eta^R = \eta^R_0 $.  

$ v_s^2(r) $ becomes {\bf large and positive} below and near $
\eta^R_0 $ and {\bf large and negative} above and near $ \eta^R_0 =
2.43450\ldots  $ as
depicted in fig. \ref{vsonr}. This singular behaviour witness the
appearance of strong instabilities at $ \eta^R = \eta^R_0 $. For 
$ \eta^R > \eta^R_0 , \; v_s(r) $ becomes imaginary indicating the
exponential growth of disturbances in the gas. This phenomenon is
especially dramatic in the denser regions (the core). 

For  $ \eta^R $ beyond $ \eta^R_0 $ and before $ \eta^R_1 =
2.14675\ldots $ in the second sheet $ v_s^2(r) $ stays negative around
the core while it becomes positive in the external regions as depicted
in fig. \ref{vsonrsup}. For example, $ v_s^2(r) $ is positive at $ \eta^R =
\eta^R_C $ for $ r > 0.4526\ldots $. 

For  $ \eta^R $ in the second sheet beyond $ \eta^R_1 $ and before $
\eta^R_{MC} , \;  v_s^2(r) $ is positive in the core and negative
outside. 

\section{$\nu$-dimensional generalization}

The  self-gravitating gas can be studied in $\nu$-dimensional space where
the Hamiltonian takes the form
\begin{equation}\label{hamiD}
H_N = \sum_{l=1}^N\;{{p_l^2}\over{2m}} - G \, m^2 \sum_{1\leq l < j\leq N}
{1 \over { |{\vec q}_l - {\vec q}_j|_A^{\nu-2}}},\quad  {\rm for}\;  \nu \neq 2
\end{equation}
and
\begin{equation}\label{hami2}
H_N = \sum_{l=1}^N\;{{p_l^2}\over{2m}} - G \, m^2 \sum_{1\leq l < j\leq N}
\log{1 \over { |{\vec q}_l - {\vec q}_j|_A}}, \quad  {\rm at}\;  \nu= 2\; .
\end{equation}

The partition function in the microcanonical, canonical and grand
canonical ensembles takes forms analogous to eqs.(II.2),
(III.1) and (VI.7) in paper I, respectively.  

We now find for the microcanonical ensemble,
$$
S(E,N)  = \log\left[ {N^{\nu N-2} \, m^{3\nu N/2-2} \,
L^{\nu(2-\nu/2)N +\nu -2} \, G^{\nu N/2 -1}
\over N !\, \Gamma\left( {\nu N \over 2} \right)\, {(2\pi)}^{\nu N/2}}\right]
+ \log w(\xi,N) \; .
$$
where the coordinate partition function takes now the form
$$
w(\xi,N)\equiv \int_0^1\ldots \int_0^1 \prod_{l=1}^N\; d^\nu r_l \; 
\left[\xi + {1 \over N}u({\vec r}_1, \ldots, {\vec r}_N) \right]^{\nu N/2
-1}\theta\left[\xi + 
{1 \over N}u({\vec r}_1, \ldots, {\vec r}_N) \right]\; .
$$
with
\begin{equation}\label{tzinu}
\xi = { E \, L^{\nu-2} \over G \, m^2 \, N^2}
\end{equation}
and
$$
u({\vec r}_1, \ldots, {\vec r}_N) \equiv {1 \over
N}\sum_{1\leq l < j\leq N} {1 \over { |{\vec r}_l - {\vec r}_j|_a^{\nu-2}}}\; .
$$

In the canonical ensemble we obtain now,
$$
{\cal Z}_C(N,T) = {1 \over N !}\left({m T
L^2\over{2\pi}}\right)^{\frac{\nu N}2}
\; \int_0^1\ldots \int_0^1
\prod_{l=1}^N d^\nu r_l\;\; e^{ \eta \; u({\vec r}_1,\ldots,{\vec r}_N)}
$$
where the variable $ \eta $ takes the form
\begin{equation}\label{etanu}
\eta = {G \, m^2 N \over L^{\nu-2} \; T} \; .
\end{equation}
As we can see from eqs.(\ref{tzinu})-(\ref{etanu}) the only change
going off three dimensional space is in the exponent of $L$.

\bigskip

In $ \nu$-dimensional space the thermodynamic limit is 
defined as $ V, \; N  \to \infty $ keeping  $ \eta $ and $ \xi $
fixed. That is, $ N/ L^{\nu-2} = N /
V^{1-2/\nu} $ is kept fixed. The volume density of particles $ N/V $
vanishes as  $ V^{-2/\nu} $ for $ V, \; N  \to \infty $ and $ \nu > 2
$. It is a  dilute limit for $ \nu > 2 $.

When $ \nu \leq  2 $, one has to assume that the temperature tends to
infinity in the thermodynamic limit in order to keep $ \eta $ and $ \xi $
fixed as $ V, \; N  \to \infty $. 

\section{The Interstellar Medium}

The interstellar medium (ISM) is a gas essentially formed by atomic (HI) 
and molecular ($H_2$) hydrogen, distributed in cold ($T \sim 5-50 K$) 
clouds, in a very inhomogeneous and fragmented structure. 
These clouds are confined in the galactic plane 
and in particular along the spiral arms. They are distributed in 
a hierarchy of structures, of observed masses from 
$10^{-2} \; M_{\odot}$ to $10^6 M_{\odot}$. The morphology and
kinematics of these structures are traced by radio astronomical 
observations of the HI hyper fine line at the wavelength of 21cm, and of
the rotational lines of the CO molecule (the fundamental line being
at 2.6mm in wavelength), and many other less abundant molecules.
  Structures have been measured directly in emission from
0.01pc to 100pc, and there is some evidence in VLBI (very long based 
interferometry) HI absorption of structures as low as $10^{-4}\; pc = 20$ AU 
(3 $10^{14}\; cm$). The mean density of structures is roughly inversely
proportional to their sizes, and vary 
between $10$ and $10^{5} \; atoms/cm^3$ (significantly above the 
mean density of the ISM which is about 
$0.1 \; atoms/cm^3$ or $1.6 \; 10^{-25}\; g/cm^3$ ).
Observations of the ISM revealed remarkable relations between the mass, 
the radius and velocity dispersion of the various regions, as first 
noticed by Larson \cite{larson}, and  since then confirmed by many other 
independent observations (see for example ref.\cite{obser}). 
From a compilation of well established samples of data for many different  
types of molecular clouds of maximum linear dimension (size) $ R $,  
total mass $M$ and internal velocity dispersion $ \Delta v$ in each region: 
\begin{equation}\label{vobser}
M (R)  \sim    R^{d_H}     \quad        ,     \quad  \Delta v \sim R^q \; ,
\end{equation}
over a large range of cloud sizes, with   $ 10^{-4}\; - \; 10^{-2}
\;  pc \;   \leq     R   \leq 100\;  pc, \;$
\begin{equation}\label{expos}
1.4    \leq   d_H    \leq   2     ,   \;     0.3  \leq     q  \leq
0.6 \; . 
\end{equation}
These {\bf scaling}  relations indicate a hierarchical structure for the 
molecular clouds which is independent of the scale over the above 
cited range; above $100$ pc in size, corresponding to giant molecular clouds,
larger structures will be destroyed by galactic shear.

These relations appear to be {\bf universal}, the exponents 
$d_H , \; q$ are almost constant over all scales of the Galaxy, and
whatever be  
the observed molecule or element. These properties of interstellar cold 
gas are supported first at all from observations (and for many different 
tracers of cloud structures: dark globules using $^{13}$CO, since the
more abundant isotopic species $^{12}$CO is highly optically thick, 
dark cloud cores using $HCN$ or $CS$ as density tracers,
 giant molecular clouds using $^{12}$CO, HI to trace more diffuse gas, 
and even cold dust emission in the far-infrared).
Nearby molecular clouds are observed to be fragmented and 
self-similar in projection over a range of scales and densities of 
at least $10^4$, and perhaps up to $10^6$.

The physical origin as well as the interpretation of the scaling relations 
 (\ref{vobser}) have been the subject of many proposals. It is not our
 aim here to account for all the proposed models  
of the ISM and we refer the reader to refs.\cite{obser} for a review.

The physics of the ISM is complex, especially when we consider the violent
perturbations brought by star formation. Energy is then poured into 
the ISM either mechanically through supernovae explosions, stellar winds,
bipolar gas flows, etc.. or radiatively through star light, heating or
ionizing the medium, directly or through heated dust. Relative velocities
between the various fragments of the ISM exceed their internal thermal
speeds, shock fronts develop and are highly dissipative; radiative cooling
is very efficient, so that globally the ISM might be considered 
isothermal on large-scales. 
Whatever the diversity of the processes, the universality of the
scaling relations suggests a common mechanism underlying the physics.

  We proposed that self-gravity is the main force at the origin of the 
structures, that can be perturbed locally by heating sources\cite{natu,prd}. 
Observations are compatible with virialised structures at all scales.
 Moreover, it has been suggested that the molecular clouds ensemble is
in isothermal equilibrium with the cosmic background radiation at $T \sim 3 K$
in the outer parts of galaxies, devoid of any star and heating
sources \cite{pcm}. This colder isothermal medium might represent the ideal
frame to understand the role of self-gravity in shaping the hierarchical
structures. 

\bigskip

In order to compare the properties of the self-gravitating gas with
the ISM it is convenient to express $ m, \; T $ and $ L $ in $ \eta $
in appropriate units. We find from eq.(\ref{defeta})
$$
\eta = 0.52193 \;  { m \; {\cal M}_{\odot}  \over L \; T} \; ,
$$
where $m$ is in multiples of the hydrogen atom mass, $T$ in Kelvin, $L$
in parsecs and  $ {\cal M}_{\odot} $ is the mass of the cloud in units
of solar masses. Notice that $ L $ is many times ($ \sim 10 $) the
size of the cloud. 

The observed parameters of the ISM clouds\cite{obser} yield an $ \eta
$ around $ \sim 2.0 $ for clouds not too large: $ {\cal M}_{\odot} 
%\lesssim 1000 $. 
< 1000 $. 
Such $ \eta $ is in the range where the
self-gravitating gas exhibits scaling behaviour. 

We conclude that the self-gravitating gas in thermal equilibrium well
describe the observed fractal structures and the scaling relations in
the ISM clouds [see, for example fig. \ref{fig12} and table 2]. Hence,
self-gravity  accounts for the structures in the ISM.

\section{Discussion}

We have presented in paper I and here a set of new results for the
self-gravitating thermal gas obtained by Monte Carlo and analytic
methods. They provide a complete picture for the thermal self-gravitating gas.

Starting from the partition function of the self-gravitating gas, we
have proved from a microscopic calculation that the local equation
of state $ p(\vec r) = T \; \rho_V(\vec r) $ and the hydrostatic
description are exact. Indeed, the dilute nature of the
thermodynamic limit ($N \sim L \to \infty $ with $N/L$ fixed) together
with the long range nature of the gravitational forces
play a crucial role in the obtention of such ideal gas equation of state.

More generally, one can investigate whether a hydrodynamical
description will apply for a self-gravitating gas. One has then to
estimate the mean free path $(l)$ for the particles and compare it
with the relevant scales $ a $ in the system \cite{llkine}. We have,
\begin{equation}\label{clm}
l \sim { 1 \over \rho_V \; \sigma_t } \sim {L^3 \over N \; \sigma_t}
\end{equation}
where $ \rho_V = {N \over L^3} \; \rho $ is the volume density of
particles and $ \sigma_t $ the total transport cross section. 

Due to the
long range nature of the gravitational force, $ \sigma_t $ diverges
logarithmically for small angles. On a finite volume the impact
parameter is bounded by $ L $ and the smaller scattering angle is of
the order of
$$
{ \Delta q \over q} \sim {G\; m^2 \over L \, T}
$$
since $ q = mv \sim \sqrt{m \; T} $ and $ \Delta q \sim {G \, m^2
\over L^2} {L \over v} \sim {G \, m^{5/2} \over L \, \sqrt{T} }$. 

We then have for the transport cross section\cite{llkine},
\begin{equation}\label{sectra}
\sigma_t \sim {(G\, m)^2 \over |\vec v - \vec v'|^4}\log{L\, T \over G\,m^2}
\sim \left({L\, N \over \eta}\right)^2 \log{N \over \eta}
\end{equation}
where we used that $ v \sim \sqrt{T \over m} $ and $ |\vec v - \vec
v'| \sim \Delta q / m $. As we see, the collisions with very large
impact parameters ($\sim L$) dominate the cross-section. 

From eqs.(\ref{clm}) and (\ref{sectra}), we find for the mean free
path: 
\begin{equation}\label{knudsen}
l  \sim {L \over N} \; \left({G\,m^2 \over T \, L}\right)^2
{1 \over \log{T \, L\over G\,m^2}} \sim {L \over N^3} {\eta^2 \over
\log{N \over \eta}} \; ,
\end{equation}
where we have here replaced $ \rho_V $ by $ {N \over L^3} $.
A more accurate estimate introduces the factor $ \rho = e^{\phi} $ in
the denominator of $ l $. This factor for spherical symmetry can vary
up to two orders of magnitude [see fig. \ref{rho0}] 
but it does not change essentially the estimate (\ref{knudsen}). 

We see from eq.(\ref{knudsen}) that  in the thermodynamic limit $ l $
becomes extremely small compared with any length $ a = {\cal O} (N^0)
$ that stays fixed for $ N \to \infty $. We find from
eq.(\ref{knudsen}),
$$
{l \over a} \sim {1 \over N^2} {\eta^2 \over \log{N \over \eta}} \; ,
$$
In conclusion, the smallness of the ratio $l/a$ (Knudsen number)
guarantees that the hydrodynamical description for a self-gravitating
fluid becomes exact in the $ N,\; L \to \infty $ limit for all scales
ranging from the order $ L^0 $ till the order $ L $.

It must me noticed that the time between two collisions $ t_{col}
   = l/v \sim l \, \sqrt{m \over T}  $ is different both from the
   relaxation time and from the  crossing time used in the
   literature. In particular, it is well known that\cite{sas,bt}
   $$
{ t_{crossing} \over t_{relaxation} } \sim {8 \over N} \, \log N \; .
$$
This formula does not concern the time $ t_{col} $ between two
   successive collisions. The time  $ t_{col} $ is indeed very short due to the
   small angle behaviour of the gravitational cross section. For
   constant cross sections one finds a very different result for $
   t_{col} $ [see ref. \cite{bt}].

\bigskip

In this paper and in its companion paper \cite{I} we thoroughly investigate the
physics of the self-gravitating gas in thermal equilibrium. It is
natural to study now the hydrodynamics of the self-gravitating fluid
using $ p(\vec r) = T \; \rho(\vec r) $ as local equation of state. A
first work on this direction is ref.\cite{seme}. 

\section{Acknowledgements}

One of us (H J de V) thanks M. Picco for useful discussions on Monte
Carlo methods. We thank S. Bouquet for useful discussions. 

\appendix

\section{Calculation of Functional Determinants in the Spherically
Symmetric Case}

We compute here the determinant of the one-dimensional differential
operator:
$$
D_1(l) = -{d^2\over dr^2} - \frac2{r} {d
\over dr} + {l(l+1) \over r^2}  -  4\pi \eta^R \; e^{\phi(r)} 
$$
 where $ \phi(r) $ is the stationary point given by
 eqs.(\ref{fixi})-(\ref{ecuaxi}). 

It is convenient to change the variable $ r $ to
$$
x \equiv \log r \quad , \quad 0 \leq r \leq 1 \quad , \quad -\infty
\leq x \leq 0 \; ,
$$
and perform a similarity transformation by $ \sqrt{r} = e^{x/2} $ on
$D_1(l)$. That is, we define
$$
D(l) \equiv e^{-x/2} \;D_1(l) \; e^{x/2} = -{d^2 \over dx^2} - k^2 +  v(x)\; ,
$$
where,
$$
v(x) \equiv -4\pi \eta^R \; e^{\phi(r = e^{x})+2x} \quad , \quad k
\equiv i \left(l+\frac12 \right) \; .
$$
$ D(l) $ has the form of a standard Schr\"odinger operator. Notice
that,
$$
\lim_{x \to -\infty } v(x) = 0\quad \mbox{and} \quad v(0) = - 3 \,
\eta^R \, f_{MF}( \eta^R) < 0 \; .
$$
We thus have an `attractive' potential $ v(x) $. The appearance of a
`bound state'  (that is, a negative eigenvalue $ k^2 < 0 $) corresponds here to
an instability in the self-gravitating gas.

We compute now the determinant
$$
\Delta_l  \equiv\Delta(k) \equiv { \mbox{Det}\left[ -{d^2 \over dx^2}
- k^2 +  v(x) \right] \over \mbox{Det}\left[ -{d^2 \over dx^2} - k^2 \right]}
$$
where we normalize at $ v = 0 $, as usual.

The logarithmic derivative with respect to $k^2$ can be expressed in
terms of the inverse of the operator $ D(l) $
\begin{equation}\label{caldet}
-{ \partial \log \Delta(k) \over \partial k^2 } = \int_{-\infty }^0 dx
\left[ \left(x\right| { 1 \over  -{d^2 \over dx^2} - k^2 +  v(x) }
\left|x \right) - \left(x\right| { 1 \over  -{d^2 \over dx^2} - k^2  }
\left|x \right)  \right]\; .
\end{equation}
In order to express the Green function (inverse of the operator $ D(l)
$) we introduce the Jost and regular solutions of the Schr\"odinger-like
equation
\begin{equation}\label{schro}
\left[ -{d^2 \over dx^2} - k^2 +  v(x)\right] f(k,x) = 0 \; .
\end{equation}
The solution of eq.(\ref{schro}) with the asymptotic behaviour
\begin{equation}\label{jost}
f(k,x) \buildrel{ x \to -\infty }\over = e^{-ikx} = e^{\left(l+\frac12
\right)x} \buildrel{ r \to 0 }\over = r^{l+\frac12} \; .
\end{equation}
is called the Jost  solution, while the solutions $ \varphi(k,x) $ of
eq.(\ref{schro}) obeying the boundary conditions
\begin{equation}\label{deffi}
\varphi(k,0) = \alpha \quad , \quad {\dot \varphi}(k,0) = \beta
\end{equation}
where $ \alpha $ and $\beta$ are $k$-independent arbitrary
parameters, are called regular solutions. The values of $ \alpha $ and
$\beta$ must be selected on physical grounds as discussed in
sec. III. $  \Delta(k) $ explicitly depends on them as we shall see below.

For large and negative $ x $ we have
$$
\varphi(k,x) \buildrel{ x \to -\infty }\over = F(k) \; e^{ikx}+G(k)\;  e^{-ikx}
$$
where the coefficients $ F(k) $ and $ G(k) $ depend on $ v(.) $. $
F(k) $ is called the Jost function. 

The small fluctuations defined by eq.(\ref{armesf}) are related to the
regular solutions by 
\begin{equation}\label{flufi}
y_l(r) = { 1 \over \sqrt{r}} \; 
\varphi(k=i \left(l+\frac12 \right), \, x=\log r) 
\end{equation}
where $ y_l(r) $ is the radial part of $ Y(\vec x) $ in the $l$th.
wave [see eq.(\ref{armesf})]. 

\medskip

The Green function of $ D(l) $ takes then the form
\begin{equation}\label{funG}
\left(x\right| { 1 \over  -{d^2 \over dx^2} - k^2 +  v(x) } \left|x'
\right)= {i \over 2\,k} \; \varphi(k,x_{>}) \;  f(k,x_{<}) \; ,
\end{equation}
where $ x_{>} $ ($ x_{<} $) stands for the larger (the smaller)
between $ x $ and $ x' $. 

Inserting eq.(\ref{funG}) into eq.(\ref{caldet}), the integration over
$ x $ can be performed with the help of wronskian identities with the result
$$
\Delta(k) = { 2 \, k \; F(k) \over \alpha \; k - i \; \beta}\; ,
$$
where we used that $ \Delta(\infty) = 1 $ as normalization. 

The Jost function $  F(k) $ can be expressed in terms of the Jost
solution at the origin just computing at the origin the wronskian of
the regular and Jost solutions,
$$
F(k) = {1 \over 2\,i\,k} \left[ \beta f(k,0) -\alpha  {\dot f}(k,0)
\right]\; .
$$
Therefore, once the Jost solution is known, the calculation of the
determinant is immediate
\begin{equation}\label{detfunj}
\Delta(k) ={\beta f(k,0) -\alpha  {\dot f}(k,0) \over \beta +i\, \alpha
\; k}
\end{equation}
Notice that $\Delta(k)$ is an homogeneous function of $ \alpha $ and $
\beta$.

We impose the physical boundary conditions (\ref{condfis})
to the regular solution (\ref{deffi}). This gives
\begin{equation}\label{alfabeta}
(2 \, l+1)\alpha + 2 \, \beta = 0
\end{equation}
and
\begin{equation}\label{deltal}
\Delta_l =\frac12 f(k,0) + {1 \over 2l+1}{\dot f}(k,0) 
\end{equation}

\begin{figure}[t] 
\begin{turn}{-90}
\epsfig{file=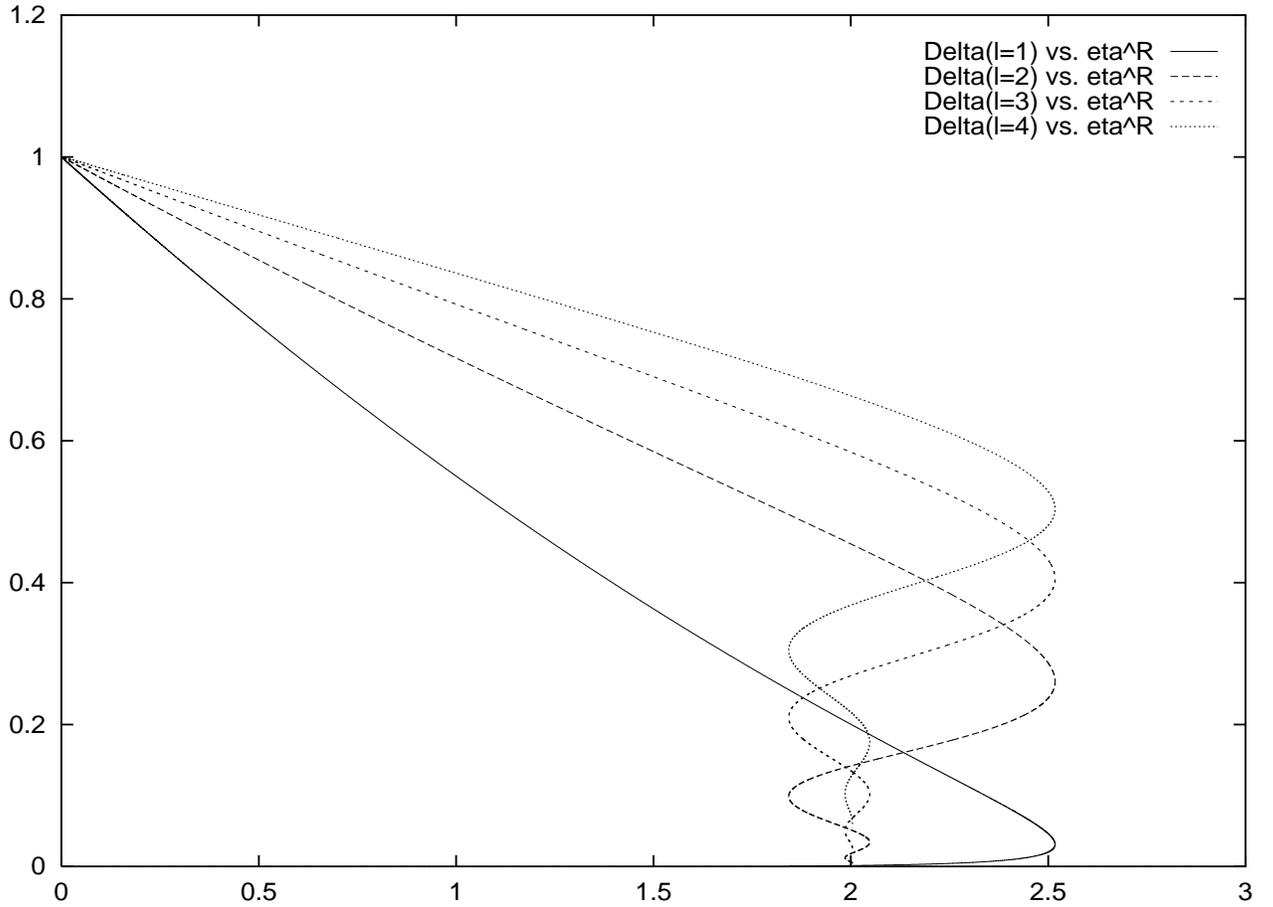,width=12cm,height=18cm} 
\end{turn}
\caption{The partial wave determinants for $l=1,\, 2,\,3 $ and $4$ as
a function of $ \eta^R $. Notice that all these determinants are
positive definite for $ \eta^R_C \geq \eta^R \geq 0$.\label{fig13}}
\end{figure}

\subsection{The S-wave determinant}

For $ l=0 , \; k = i/2 $, eq.(\ref{schro})  can be solved in terms of the
stationary point solution $ \phi(r = e^{x}) $. That is,
\begin{equation}\label{ondaS}
f(i/2,x) = e^{x/2} \left[ 1 + \frac12 \; {d \phi(r = e^{x}) \over dx}
\right] \; ,
\end{equation}
obeys both eq.(\ref{schro}) and the boundary conditions (\ref{jost}). 
Eq.(\ref{schro}) for the function (\ref{ondaS}) can be checked just taking the
derivative of the stationary point equation (VI.43) in paper I
with respect to $ x = \log r $. 

We can now compute $ f(i/2,0) $ and $ {\dot f}(i/2,0) $ using
eqs.(VI.43), (VI.46) in paper I, and  (\ref{fide1}) with the
result,
$$
f(i/2,0) = 1 -{\eta^R \over 2}\quad , \quad  {\dot f}(i/2,0)= \frac12
+ {\eta^R \over 4}-\frac32 \; \eta^R \; f_{MF}(\eta^R) \; .
$$
This yields for the S-wave determinant (\ref{detfunj}) 
$$
\Delta_0 = \Delta(k=i/2) = { 1 \over \alpha - 2 \, \beta } \, 
\left\{  \alpha \left[ 1 +{\eta^R \over 2}-3\, \eta^R \;
f_{MF}(\eta^R) \right] + \beta\; (\eta^R -2) \right\}
$$
for arbitrary values of $ \alpha $  and $ \beta $.

For the boundary conditions (\ref{alfabeta}) we find
\begin{equation}\label{detS}
\Delta_0(\eta^R) = 1 - \frac32 \, \eta^R \, f_{MF}(\eta^R) \; .
\end{equation}

\subsection{The P-wave determinant}

For $ l=1 , \; k = 3 i/2 $, eq.(\ref{schro})  can also be solved in
terms of the stationary point solution $ \phi(r = e^{x}) $. That is,
\begin{equation}\label{ondaP}
f(3i/2,x) = -{3 \over \lambda^2}\; e^{-x/2} \; {d \phi(r = e^{x}) \over dx}
\end{equation}
obeys both eq.(\ref{schro}) and the boundary conditions (\ref{jost}). 
Eq.(\ref{schro}) for the function (\ref{ondaP}) can be checked just taking the
derivative with respect to $ r $ of the stationary point equation
(VI.43) in paper I.

We obtain for $ f(3i/2,0) $ and $ {\dot f}(3i/2,0) $,
$$
f(3i/2,0) = { 3 \, \eta^R \over  \lambda^2} \quad , \quad 
{\dot f}(3i/2,0) = { 9 \over  \lambda^2} \left[ f_{MF}(\eta^R) -
{\eta^R \over 2}\right] 
$$
where we used again eqs.(VI.43), (VI.46) in paper I,  and
(\ref{fide1}).

We finally obtain for the P-wave determinant for arbitrary
values of $ \alpha $  and $ \beta $,
$$
\Delta_1=\Delta(k=3i/2) = { 3 \, \eta^R \over  \lambda^2 (3 \, \alpha
- 2 \, \beta )} \left\{ 3 \, \alpha \left[ 2 \, f_{MF}(\eta^R)  -1
\right] - 2 \, \beta \right\} 
$$
For the boundary conditions (\ref{alfabeta}) we find
\begin{equation}\label{detP}
\Delta_1(\eta^R) = { 3 \, \eta^R \over \lambda^2}f_{MF}(\eta^R) \; .
\end{equation}

\subsection{The D-wave and higher waves}

We plot in fig. \ref{fig13} the determinants for $ l = 2, 3, 4 $ and $ 5 $ for
the boundary conditions (\ref{alfabeta}) We see that these partial waves
determinants are positive for all positive values of $ \eta^R \leq \eta^R_C $.

We can evaluate the Jost solutions asymptotically for large $ k $
(large $l$). Using the standard Riccati transformations yields
$$
f(k,x)= e^{(l+\frac12 )x} \left[ 1 + { I(x) \over 2 l + 1 } -{ I^2(x)
+ 2 \, v(x) \over 2(2 l + 1)^2 } + {\cal O}\left( {1 \over (2 l +
1)^3}  \right) \right]
$$
where $ k = i \left(l+\frac12 \right) $ and
$$
I(x) \equiv \int_{-\infty}^x dy \, v(y) = \lambda \, e^x \,
\chi'(\lambda e^x) +\chi(\lambda e^x)
$$

Therefore, for the boundary conditions (\ref{alfabeta}) we find from
eq.(\ref{deltal}), 
$$
\Delta_l(\eta^R) = 1 + { I(0) \over 2 l + 1 } - { I^2(0) \over 2(2 l +
1)^2 } + {\cal O}\left( {1 \over (2 l +
1)^3}  \right) 
$$
where,
$$
I(0) = \log\left[{ 3 \, \eta^R \over \lambda^2}f_{MF}(\eta^R)\right] -
\eta^R \; .
$$
\section{Calculation of $<r>$ and $<r^2>$ in the mean field}

In the mean field approach we have from eqs.(\ref{defryr2}) and (\ref{correMF})
$$
<r> = \int_0^1 r^2 \; dr  \int_0^1 {r'}^2 \; dr' \; \int d\Omega(\hat r) 
\; \int d\Omega(\hat r') \; \rho(r) \;\rho(r') \; \sqrt{r^2+{r'}^2 - 2 r
r' \; \cos({\hat r, \hat r'})} \; ,
$$
$$
<r^2> = \int_0^1 r^2 \; dr  \int_0^1 {r'}^2 \; dr' \; \int d\Omega(\hat r) 
\; \int d\Omega(\hat r') \; \rho(r) \;\rho(r') \; [r^2+{r'}^2 - 2 r
r' \; \cos({\hat r, \hat r'})] \; .
$$
Integrating over the angles yield,
$$
<r> = {16\, \pi^2 \over 3} \int_0^1 r \, dr \;  \rho(r) \left\{
\int_0^r {r'}^2 \; dr' \;\rho(r') \left[ {r'}^2 + 3 \, r^2 \right] + r \;
\int_r^1 r' \; dr' \; \rho(r') \left[ r^2 + 3 \, {r'}^2 \right] \right\}
$$
$$
<r^2> = 8\pi \; \int_0^1 r^4 \; dr \; \rho(r)
$$
We have then to compute integrals of the type $ \int r^n \; dr \; 
e^{\phi(r)} \; $ for $ n=1, \; 2, \; 3 $ and $ 4 $. We find using
eqs.(VI.43), (VI.45), (VI.46) and (VI.54) in paper I,
\begin{eqnarray}
4 \, \pi \, \eta^R \int_0^r {r'}^2 \; e^{\phi(r')} \; dr' &=& - r^2 \; {d
\phi(r) \over dr} \cr \cr
4 \, \pi \, \eta^R \int_r^1 r' \; e^{\phi(r')} \; dr' &=& {d
 \over dr}[r \; \phi(r)] + \eta^R - \phi(1)  \cr \cr
4 \, \pi \, \eta^R \int_r^1 {r'}^3 \; e^{\phi(r')} \; dr' &=& r^3 \, {d\phi(r)
 \over dr} - r^2 \phi(r) + \eta^R + \phi(1)  - 2 \int_r^1 r' \;
\phi(r') \; dr' \cr \cr
4 \, \pi \, \eta^R \int_0^r {r'}^4 \; e^{\phi(r')} \; dr' &=& -r^4 \, {d\phi(r)
 \over dr} +2\, r^3 \phi(r) - 6 \,  \int_0^r {r'}^2 \;
\phi(r') \; dr' 
\end{eqnarray}
Collecting all terms yields after calculation eq.(\ref{ryr2exp}).

\end{document}